# Thermodynamics and kinetics of nucleation in binary solutions

Nikolay V. Alekseechkin

Akhiezer Institute for Theoretical Physics, National Science Centre "Kharkov Institute of Physics and Technology", Akademicheskaya Street 1, Kharkov 61108, Ukraine
Email: n.alex@kipt.kharkov.ua

A new approach that is a combination of classical thermodynamics and macroscopic kinetics is offered for studying the nucleation kinetics in condensed binary solutions. The theory covers the separation of liquid and solid solutions proceeding along the nucleation mechanism, as well as liquid-solid transformations, e.g., the crystallization of molten alloys. The cases of nucleation of both unary and binary precipitates are considered. Equations of equilibrium for a critical nucleus are derived and then employed in the macroscopic equations of nucleus growth; the steady state nucleation rate is calculated with the use of these equations. The present approach can be applied to the general case of non-ideal solution; the calculations are performed on the model of regular solution within the classical nucleation theory (CNT) approximation implying the bulk properties of a nucleus and constant surface tension. The way of extending the theory beyond the CNT approximation is shown in the framework of the finite-thickness layer method. From equations of equilibrium of a surface layer with coexisting bulk phases, equations for adsorption and the dependences of surface tension on temperature, radius, and composition are derived. Surface effects on the thermodynamics and kinetics of nucleation are discussed.

## 1. Introduction

Binary nucleation covers a wide class of processes of phase transformations which can be divided into three groups: (i) gas-liquid (or solid) transformations, (ii) liquid-gas transformations, and (iii) transformations within a condensed state. The first group includes the binary droplet nucleation in a mixture of vapors of two substances [1-4], whereas the second group involves the bubble nucleation in binary fluids [5-7]. The third group includes liquid-liquid (LL), solid-solid (SS), and liquid-solid (LS) transformations. LL and SS transformations are the separation of liquid and solid solutions. LS transformation is the crystallization of a liquid alloy as well as the precipitation of a dissolved substance from a supersaturated liquid solution. Such a division is due to the different physics of the nucleation process within these groups, i.e. different equations of equilibrium for a critical nucleus as well as growth equations for a postcritical one. These equations are common for the processes of the third group, so that LL, SS, and LS binary nucleation is the subject of the present theory. As a particular case, the nucleation of one-component precipitates from a binary solution is also considered.

Binary as well as multicomponent nucleation is described in the framework of the formalism of the multivariable theory of nucleation [8] which is a universal theory – it describes the nucleation processes in different systems according to the same algorithm [1, 9-11]. Taking into account any phenomenon leads to the appearance of the corresponding variable in the theory and thereby the accuracy of the process description is increased. E.g., the addition of droplet temperature to the theory of vapor condensation and the consideration of heat exchange between the droplet and vapor allow us to calculate nonisothermal effects in nucleation [1, 11]. Taking into account surface effects [12, 13] essentially advances the theory beyond the CNT approximation.

The classical work by Reiss [14] on binary nucleation can be considered also as the first work on the multivariable theory of nucleation; the basic concepts of the latter where introduced therein – the flux of nuclei in the phase space and the saddle surface representing the work of nucleus formation. The physical picture of nucleation was shown as the flow of nuclei in the saddle surface "gorge" through its pass. In the multivariable theory, a new-phase nucleus is described by the set of variables $\{X_i\}$ one of which, $X_1$, is "unstable" ; it describes the nucleus size - $X_1 = V$, the nucleus volume, $X_1 = R$, the radius, or $X_1 = N$ the total number of particles. The remaining variables $X_2,...,X_n$ are "stable", so that the work of nucleus formation $W(X_1,...,X_n)$ is represented by a saddle surface in the $n$-dimensional space. In the vicinity of the saddle point $\{X_i^*\}$, it can be expanded up to quadratic terms:

$$W(X_1,...,X_n) = W_* + (1/2)\sum_{i,k=1}^{n} h_{ik}(X_i - X_i^*)(X_k - X_k^*) \tag{1}$$

Reiss' work being an extension of the Zeldovich-Frenkel [15, 16] one-dimensional theory to binary case employs the *microscopic* kinetics; it operates with the probabilities $w_i^{(+)}$ and $w_i^{(-)}$ of attachment and detachment of each kind monomers. This approach has become traditional for the binary-nucleation theory [17-21]; in particular, its finite difference equations are convenient for numerical studies of nucleation [3, 4, 18, 19]. In contrast to it, the approach of *macroscopic* kinetics is used in the present theory, as in the previous works [1, 9-12]. The advantages of this approach from the physical and analytical points of view were shown in Ref. [1] by the example of binary droplet nucleation. In particular, it is natural to use the fractions $x_i = N_i / N$ as the variables of nucleus description [8, 22], rather than the numbers $N_i$; the basic equations of the theory – the equations of equilibrium and growth equations – are formulated just in terms of $x_i$, as shown below.

The basis of the offered approach is the equations of motion of a nucleus in the space $\{X_i\}$. In the vicinity of the saddle point, they are linear [8]:

$$\dot{X}_i = -\sum_{k=1}^{n} z_{ik}(X_k - X_k^*), \quad \mathbf{Z} = \mathbf{BH}/kT \tag{2}$$

where $\dot{X}_i \equiv dX_i/dt$; $\mathbf{B}$ is the matrix of "diffusivities" in the Fokker-Planck equation for the distribution function of nuclei $F(X_1, ..., X_n; t)$. These equations are obtained from the conditions that the flux of nuclei

$$J_i = -b_{ij}\frac{\partial F}{\partial X_j} + \dot{X}_i F \tag{3}$$

is equal to zero for the equilibrium distribution function $F_e(X_1, ..., X_n) \sim \exp(-W(\{X_i\})/kT)$. As is known from the theory of the Fokker-Planck equation,

$$\dot{X}_i = \lim_{\Delta t \to 0} \frac{\langle \Delta X_i \rangle_{\Delta t}}{\Delta t} \tag{4}$$

where the averaging over all possible changes $\Delta X_i$ (with the corresponding probability) in the time $\Delta t$ is done. Thus, despite the fact that the actual motion of a nucleus in the vicinity of the saddle point is chaotic (Brownian), only the *regular* component of this motion remains after the procedure of averaging, i.e. the velocities $\dot{X}_i$ are *macroscopic*.

An equation for the steady state nucleation rate was derived in Ref. [8]:

$$I = C_0 \sqrt{\frac{kT}{2\pi}|h_{11}^{-1}|} |\kappa_1| e^{-\frac{W_*}{kT}} \tag{5}$$

where $h_{11}^{-1}$ is the matrix $\mathbf{H}^{-1}$ element; $\kappa_1$ is the negative eigenvalue of the matrix $\mathbf{Z}$, and $C_0$ is the normalizing factor of the *one-dimensional* equilibrium distribution function $F_e(X_1)$ - it is determined in the framework of statistical mechanical approach. This equation essentially corrects the preceding result by Trinkaus [23]; it is invariant with respect to the space dimensionality and gives the result of the Zeldovich-Frenkel theory for $n = 1$. Also, the derivation of this equation in Ref. [8] does not employ the simultaneous diagonalization of the matrices $\mathbf{H}$ and $\mathbf{B}$, which is possible only if the matrix $\mathbf{B}$ is symmetric; therefore, Eq. (5) is applicable also to the cases, when the matrix $\mathbf{B}$ includes antisymmetric elements [10].

The work by Russell [24] on nucleation in condensed phases should be also mentioned. The nucleation of a binary precipitate of a *fixed* composition is studied therein within the "shell model"; the variables of the theory are $N$, number of A atoms in the nucleus and $x$, number of A atoms in the nucleus' shell of nearest neighbors. In other words, the variable $N$ relates to the nucleus, whereas $x$ does not belong to it. As is seen from the above description of a multivariable theory, this model does not correspond to it. Therefore, Russell's model and the corresponding two-dimensionality seem artificial; the actual two-dimensionality appears, when both the variables describe the nucleus - $N$ is the total number of atoms and $x$ is the composition. When the fluctuations of nucleus composition are allowed, the problem becomes two-dimensional. The concentration of A atoms in the nucleus' shell of nearest neighbors is indeed an important quantity, however, it is considered within the present approach in connection with nucleus growth.

So, the LL, SS, and LS binary nucleation is studied here for the first time within the macroscopic approach. As is seen from the foregoing, the matrices $\mathbf{H}$ and $\mathbf{Z}$ are all we need to calculate the nucleation rate $I$. The consideration is carried out within the CNT approximation [12]: the nucleus properties are assumed the same as the properties of the bulk phase (the actual nucleus inhomogeneity and the surface effects are not taken into account). Accordingly, the surface tension $\sigma$ is constant; with the same accuracy, the partial molecular volumes $v_i$ are also constant. This approximation is justified as the first step in constructing the theory of binary nucleation in a condensed state; the aim of the paper is to show how the offered approach works in the given case. It should be noted that just this approximation is employed in the most of works on binary nucleation.

Nevertheless, the extension of the theory beyond the CNT approximation is also considered in Appendix, where the surface effects in binary nucleation are taken into account within the classical thermodynamics. Equations for adsorption and the dependences of surface tension on radius, temperature, and composition are derived. It is shown that all these dependences of surface tension are due to the nucleus inhomogeneity [12, 13]; in particular, the dependence of surface tension on composition is due to adsorption (the difference in the compositions of surface layer and bulk new phase) and therefore it makes no sense to consider this dependence within the CNT approximation, where a nucleus is homogeneous and there is no adsorption.

The outline of the paper is as follows. The thermodynamics of nucleation is considered in Section 2. The equations of equilibrium of a critical nucleus with the mother phase are derived, from which the critical radius and composition can be found. Section 3 is the kinetic part of the work: the macroscopic equations of diffusion growth of a one- and two-component nucleus are obtained here; the results of Section 2 are essentially used in these equations. In Section 4, the matrix $\mathbf{Z}$ and the nucleation rate are calculated for all cases considering in Section 2 as well as the obtained results are discussed. The summary of results is given in Section 5. In Appendix, the thermodynamics of surface layer is considered and equations for adsorption and the mentioned dependences of surface tension are derived; also, surface effects on the thermodynamics and kinetics of binary nucleation are discussed.

## 2. Equations of equilibrium for a critical nucleus

The new phase is denoted by $\alpha$, the mother phase is $\beta$. The nucleation process is considered at constant temperature $T$ and pressure $P^{\beta}$; the solution is supersaturated with respect to composition $x^{\beta}$ (the component A fraction). The Laplace equation and its differential form are essentially used below:

$$P^\alpha - P^\beta = \frac{2\sigma}{R_*} \equiv P_L, \quad dP^\alpha = dP_L \qquad (6)$$

in view of $P^\beta = const$, $dP^\beta = 0$; asterisk denotes the critical value (it is omitted in equations of equilibrium for brevity).

## 2.1. Nucleation of a one-component precipitate from binary solution

Let component A precipitates. The chemical potential of component A in a non-ideal solution is

$$\mu^\beta(T, P^\beta, x^\beta) = \bar{\mu}^\beta(T, P^\beta) + kT \ln f^\beta(T, P^\beta, x^\beta) x^\beta \qquad (7)$$

where $x^\beta$ and $f^\beta(T, P^\beta, x^\beta)$ are the fraction and the activity of component A in the solution (the subscript A is omitted for brevity); the bar relates to pure component A. The condition of equilibrium $\mu^\alpha(T, P^\alpha) = \mu^\beta(T, P^\beta, x^\beta)$ of the precipitate with the mother phase in the differential form

$$d\mu^\alpha = d\mu^\beta \qquad (8)$$

at constant $T$ and $P^\beta$ has the following form, in view of Eq. (6):

$$\upsilon^\alpha dP_L = \dot{\mu}^\beta dx^\beta, \quad \dot{\mu}^\beta \equiv \left(\frac{\partial \mu^\beta}{\partial x^\beta}\right)_{T, P^\beta} \qquad (9)$$

Eq. (8) means that the state of the system "nucleus + mother phase" changes while maintaining the equilibrium between them, i.e. the critical radius $R_*$ is adjusted to composition $x^\beta$ or vice versa. In other words, Eq. (9) is an equation for the dependence $R_*(x^\beta)$. Integrating it from $P_L = 0$ ($R_* = \infty$) to the current $P_L$, we get $\mu^\beta(x^\beta) - \mu(x_\infty^\beta) = \upsilon^\alpha P_L$, from where, in view of Eq. (7),

$$f^\beta(x^\beta) x^\beta = f^\beta(x_\infty^\beta) x_\infty^\beta \exp\left(\frac{\upsilon^\alpha P_L}{kT}\right) \qquad (10)$$

where $x_\infty^\beta(T)$ is the composition of solution over the planar interface. This equation gives the desired dependence $R_*(x^\beta)$:

$$R_*(x^\beta) = \frac{2\upsilon^\alpha \sigma}{kT} \left\{\ln \frac{x^\beta}{x_\infty^\beta} + \ln \frac{f^\beta(x^\beta)}{f^\beta(x_\infty^\beta)}\right\}^{-1} \qquad (11)$$

The quantity $S = x^\beta / x_\infty^\beta$ is the supersaturation ratio; $R_* = \infty$ for $S = 1$.

For a regular solution,

$$f^\beta(x^\beta) = \exp\left[\omega^\beta (1 - x^\beta)^2\right] \qquad (12)$$

where $\omega^\beta$ is the characteristic parameter of the solution. Thus, Eq. (10) becomes

$$x^\beta \exp\left[\omega^\beta (1 - x^\beta)^2\right] = C \exp\left(\frac{\upsilon^\alpha P_L}{kT}\right), \quad C(T) \equiv x_\infty^\beta(T) \exp\left[\omega^\beta (1 - x_\infty^\beta(T))^2\right] \qquad (13)$$

from where

$$R_*(x^\beta) = \frac{2\upsilon^\alpha \sigma}{kT} \left\{ \ln \frac{x^\beta}{x_\infty^\beta} + \omega^\beta \left[ (1-x^\beta)^2 - (1-x_\infty^\beta)^2 \right] \right\}^{-1} \tag{14}$$

The ideal solution approximation is obtained by putting $\omega^\beta = 0$, or $f^\beta = 1$; hence,

$$x^\beta = x_\infty^\beta \exp\left( \frac{\upsilon^\alpha P_L}{kT} \right) \tag{15}$$

The same equation holds for a dilute solution. This is the familiar Ostwald-Freindlich equation; it gives

$$R_*(x^\beta) = \frac{2\upsilon^\alpha \sigma}{kT} \left( \ln \frac{x^\beta}{x_\infty^\beta} \right)^{-1} \tag{16}$$

It should be noted that Eqs. (15) and (16) are the same as the corresponding equations for a droplet in vapor [11] with replacing $x^\beta$ by the vapor pressure; Eq. (15) is an analogue of the Kelvin equation for the equilibrium vapor pressure over the droplet of radius $R$.

In literature, the linearized form of Eq. (15), $x^\beta = x_\infty^\beta (1 + 2\upsilon^\alpha \sigma / R_* kT)$ is often used [25], which is valid only for large nuclei (near the binodal), or small supersaturations. At the same time, if the surface tension is not too small, the nucleation begins with nanosized nuclei and Eq. (15) has to be employed. On the other hand, Eq. (15) with constant surface tension $\sigma$ is not suitable for too small nuclei. It is seen that $x^\beta$ increases with decreasing $R$, however, it has not to exceed unity; a stronger condition $x^\beta << 1$ must be satisfied for a dilute solution. This shortcoming of CNT Eq. (15) is easily corrected by using the radius-dependent surface tension $\sigma(R)_{T,P^\beta}$ in it; $\sigma$ is radius-dependent in Eq. (6) [26] and the above procedure of integrating Eq. (9) does not imply the constancy of $\sigma$. In this way, the classical Kelvin equation was extended to small radii [13]. The linear asymptotics $\sigma(R) = K(T)R$ at $R \to 0$ [27, 28] ensures a finite (spinodal) value of $x^\beta$ in this limit, $x_s^\beta = x_\infty^\beta \exp\left( 2\upsilon^\alpha K / kT \right)$. Vice versa, the constant $K$ can be found from this equation, if the spinodal value $x_s^\beta$ is known.

## 2.1. Nucleation of a compound from binary solution

Let the compound $A_nB_m$ precipitates from a non-ideal solution of components A and B:

$$\mu_A^\beta(T, P^\beta, x^\beta) = \overline{\mu}_A^\beta(T, P^\beta) + kT \ln f_A^\beta(T, P^\beta, x^\beta) x^\beta$$

$$\mu_B^\beta(T, P^\beta, x^\beta) = \overline{\mu}_B^\beta(T, P^\beta) + kT \ln f_B^\beta(T, P^\beta, x^\beta)(1-x^\beta) \tag{17}$$

This is the chemical reaction

$$nA + mB = A_nB_m \equiv C \tag{18}$$

in the *two-phase* system. The condition of equilibrium for chemical potentials is obtained from Eq. (18) by replacing the symbols A, B, and C by the corresponding chemical potentials [29]:

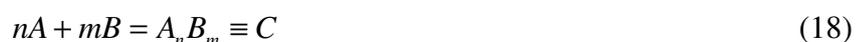

$$\mu_C^\alpha(T, P^\alpha) = n\mu_A^\beta + m\mu_B^\beta \tag{19}$$

The differential form of this equation, $d\mu_C^\alpha = n d\mu_A^\beta + m d\mu_B^\beta$, at constant $T$, $P^\beta$ and in view of Eq. (6) is

$$\upsilon_C^\alpha dP_L = n\dot\mu_A^\beta dx^\beta + m\dot\mu_B^\beta dx^\beta \tag{20}$$

where the point denotes the derivative with respect to $x^\beta$ and $\upsilon_C^\alpha = n\upsilon_A^\alpha + m\upsilon_B^\alpha$. Integration of this equation gives

$$\upsilon_C^\alpha P_L = n\left[\mu_A^\beta(x^\beta) - \mu_A^\beta(x_\infty^\beta)\right] + m\left[\mu_B^\beta(x^\beta) - \mu_B^\beta(x_\infty^\beta)\right] \tag{21}$$

where $x_\infty^\beta(T)$ is the equilibrium composition of solution over the planar interface with the compound. Employing Eq. (17), we get

$$\frac{\upsilon_C^\alpha P_L}{kT} = \ln\left\{\left(\frac{x^\beta}{x_\infty^\beta}\right)^n\left(\frac{1-x^\beta}{1-x_\infty^\beta}\right)^m\right\} + \ln\left\{\left(\frac{f_A^\beta(x^\beta)}{f_A^\beta(x_\infty^\beta)}\right)^n\left(\frac{f_B^\beta(x^\beta)}{f_B^\beta(x_\infty^\beta)}\right)^m\right\} \equiv \eta(x^\beta) \tag{22}$$

from where the critical radius of compound nucleus for the given solution composition is found as

$$R_*(x^\beta) = \frac{2\upsilon^\alpha\sigma}{kT}\left[\eta(x^\beta)\right]^{-1} \tag{23}$$

Eq. (22) is transformed to the following one:

$$(x^\beta)^n(1-x^\beta)^m\left[f_A^\beta(x^\beta)\right]^n\left[f_B^\beta(x^\beta)\right]^m = C\exp\left(\frac{\upsilon_C^\alpha P_L}{kT}\right)$$

$$C(T) \equiv (x_\infty^\beta)^n(1-x_\infty^\beta)^m\left[f_A^\beta(x_\infty^\beta)\right]^n\left[f_B^\beta(x_\infty^\beta)\right]^m \tag{24}$$

For ideal or dilute solutions, $f_A^\beta = f_B^\beta = 1$, so that only the first summand in Eq. (22) remains. For a regular solution,

$$f_A^\beta(x^\beta) = \exp\left[\omega^\beta(1-x^\beta)^2\right], \quad f_B^\beta(x^\beta) = \exp\left[\omega^\beta(x^\beta)^2\right] \tag{25}$$

and Eq. (24) acquires the following explicit form:

$$(x^\beta)^n(1-x^\beta)^m\exp\left\{\omega^\beta\left[n(1-x^\beta)^2 + m(x^\beta)^2\right]\right\} = C\exp\left(\frac{\upsilon_C^\alpha P_L}{kT}\right),$$

$$C(T) \equiv (x_\infty^\beta)^n(1-x_\infty^\beta)^m\exp\left\{\omega^\beta\left[n(1-x_\infty^\beta)^2 + m(x_\infty^\beta)^2\right]\right\} \tag{26}$$

### 2.3. Nucleation of a two-component precipitate

Differently from the previous case, here the nucleus composition is not fixed: first, the critical nucleus composition depends on its radius, $x_*^\alpha(R_*)$; second, the composition can fluctuate around the critical value $x_*^\alpha$, which makes the problem *two-dimensional*. The variance of this fluctuation will be given later. As in the previous cases, it makes sense to start with a non-ideal solution, since the activities are experimentally determined quantity. For the convenience of using the resulting equations of equilibrium in the subsequent kinetic equations, the symmetric form of thermodynamic equations with respect to both components is employed here; in particular, the fractions $x_A$ and $x_B$ are used:

$$\mu_i^\beta(T, P^\beta, x_i^\beta) = \overline{\mu}_i^\beta(T, P^\beta) + kT \ln f_i^\beta(T, P^\beta, x_i^\beta) x_i^\beta$$

$$\mu_i^\alpha(T, P^\alpha, x_i^\alpha) = \overline{\mu}_i^\alpha(T, P^\alpha) + kT \ln f_i^\alpha(T, P^\alpha, x_i^\alpha) x_i^\alpha, \quad i = \text{A, B} \tag{27}$$

The equation of equilibrium $d\mu_i^\alpha(T, P^\alpha, x_i^\alpha) = d\mu_i^\beta(T, P^\beta, x_i^\beta)$ at constant $T$, $P^\beta$ and in view of Eq. (6) gives

$$\upsilon_i^\alpha dP_L + \dot{\mu}_i^\alpha dx_i^\alpha - \dot{\mu}_i^\beta dx_i^\beta = 0, \quad \dot{\mu}_i^\alpha \equiv \left(\frac{\partial \mu_i^\alpha}{\partial x_i^\alpha}\right)_{T, P^\alpha}, \quad \dot{\mu}_i^\beta \equiv \left(\frac{\partial \mu_i^\beta}{\partial x_i^\beta}\right)_{T, P^\beta} \tag{28}$$

This is a Pfaffian equation [1] for the vector field $\mathbf{F} = \upsilon_i^\alpha \mathbf{i} + \dot{\mu}_i^\alpha \mathbf{j} - \dot{\mu}_i^\beta \mathbf{k}$ in the space $\mathbf{r} = (P_L, x_i^\alpha, x_i^\beta)$. The necessary and sufficient condition of its integrability by one relation $x_i^\beta = x_i^\beta(x_i^\alpha, P_L)$ is $\mathbf{F}(\nabla \times \mathbf{F}) = 0$. We have

$$\nabla \times \mathbf{F} = -\left(\frac{\partial \dot{\mu}_i^\beta}{\partial x_i^\alpha} + \frac{\partial \dot{\mu}_i^\alpha}{\partial x_i^\beta}\right)\mathbf{i} - \left(\frac{\partial \dot{\mu}_i^\beta}{\partial P_L} + \frac{\partial \upsilon_i^\alpha}{\partial x_i^\beta}\right)\mathbf{j} + \left(\frac{\partial \dot{\mu}_i^\alpha}{\partial P_L} - \frac{\partial \upsilon_i^\alpha}{\partial x_i^\alpha}\right) = 0$$

in view of the obvious equality to zero of the derivatives in the first two summands and

$$\frac{\partial \dot{\mu}_i^\alpha}{\partial P_L} = \frac{\partial}{\partial x_i^\alpha}\left(\frac{\partial \mu_i^\alpha}{\partial P_L}\right) = \frac{\partial \upsilon_i^\alpha}{\partial x_i^\alpha}$$

Thus, the vector field is potential, $\mathbf{F} = \nabla U$, and Eq. (28) has the form $dU = 0$; its solution is $U = const$, the integration constant is determined by the chosen initial condition.

For component A, we integrate Eq. (28) in the space ($P_L$, $x_A^\alpha$, $x_A^\beta$) along the broken line whose segments are parallel to the coordinate axes, starting from the point corresponding to the planar interface with pure component A in phase $\alpha$: ($P_L = 0$, $x_A^\alpha = 1$, $x_A^\beta = x_{A,\infty}^\beta$), where $x_{A,\infty}^\beta(T)$ is the fraction of component A in phase $\beta$ over the planar interface of pure A in phase $\alpha$. So, the path of integration is as follows: (i) $P_L$ changes from 0 to the current $P_L$ at $x_A^\alpha = 1$ and $x_A^\beta = x_{A,\infty}^\beta$; (ii) $x_A^\alpha$ changes from 1 to the current $x_A^\alpha$ at the given $P_L$ and $x_A^\beta = x_{A,\infty}^\beta$; (iii) $x_A^\beta$ changes from $x_{A,\infty}^\beta$ to the current $x_A^\beta$ at the given $P_L$ and $x_A^\alpha$. As a result, one obtains

$$\overline{\upsilon}_A^\alpha P_L + \left[\mu_A^\alpha(x_A^\alpha) - \overline{\mu}_A^\alpha\right] - \left[\mu_A^\beta(x_A^\beta) - \mu_A^\beta(x_{A,\infty}^\beta)\right] = 0 \tag{29a}$$

For component B, we integrate Eq. (28) in the space ($P_L$, $x_B^\alpha$, $x_B^\beta$) in the similar way; the starting point is ($P_L = 0$, $x_B^\alpha = 1$, $x_B^\beta = x_{B,\infty}^\beta$), where $x_{B,\infty}^\beta(T)$ is the fraction of component B in phase $\beta$ over the planar interface of pure B in phase $\alpha$. As a result,

$$\overline{\upsilon}_B^\alpha P_L + \left[\mu_B^\alpha(x_B^\alpha) - \overline{\mu}_B^\alpha\right] - \left[\mu_B^\beta(x_B^\beta) - \mu_B^\beta(x_{B,\infty}^\beta)\right] = 0 \tag{29b}$$

It should be emphasized that these equations contain the partial volumes $\overline{\upsilon}_A^\alpha$ and $\overline{\upsilon}_B^\alpha$ of *pure* components in phase $\alpha$ (which are therefore specific volumes $\upsilon_i^\alpha$), even if these quantities are composition-dependent, as a consequence of the integration path (integration over $P_L$ is performed at $x_i^\alpha = 1$).

With the use of Eq. (27), Eqs. (29a, b) become as follows:

$$\begin{cases} \ln \dfrac{f_A^\beta(x_A^\beta)x_A^\beta}{f_{A,\infty}^\beta x_{A,\infty}^\beta} - \ln f_A^\alpha(P^\alpha, x_A^\alpha)x_A^\alpha = \dfrac{\upsilon_A^\alpha P_L}{kT} \\[3mm] \ln \dfrac{f_B^\beta(x_B^\beta)x_B^\beta}{f_{B,\infty}^\beta x_{B,\infty}^\beta} - \ln f_B^\alpha(P^\alpha, x_B^\alpha)x_B^\alpha = \dfrac{\upsilon_B^\alpha P_L}{kT} \end{cases} \tag{30}$$

where $f_{A,\infty}^\beta \equiv f_A^\beta(x_{A,\infty})$, $f_{B,\infty}^\beta \equiv f_B^\beta(x_{B,\infty})$, and $P^\alpha = P^\beta + P_L$.

For a regular solution, $f_i^\alpha = \exp\!\left[\omega^\alpha(1-x_i^\alpha)^2\right]$ and $f_i^\beta = \exp\!\left[\omega^\beta(1-x_i^\beta)^2\right]$, so that Eq. (30) has the following form:

$$\begin{cases} x_A^\beta \exp\!\left[\omega^\beta(1-x_A^\beta)^2\right] = C_A x_A^\alpha \exp\!\left[\dfrac{\upsilon_A^\alpha P_L}{kT} + \omega^\alpha(1-x_A^\alpha)^2\right], \quad C_A(T) \equiv x_{A,\infty}^\beta \exp\!\left[\omega^\beta(1-x_{A,\infty}^\beta)^2\right] \\[3mm] x_B^\beta \exp\!\left[\omega^\beta(1-x_B^\beta)^2\right] = C_B x_B^\alpha \exp\!\left[\dfrac{\upsilon_B^\alpha P_L}{kT} + \omega^\alpha(1-x_B^\alpha)^2\right], \quad C_B(T) \equiv x_{B,\infty}^\beta \exp\!\left[\omega^\beta(1-x_{B,\infty}^\beta)^2\right] \end{cases} \tag{31}$$

Eqs. (30) and (31) give the desired dependence $x_i^\beta = x_i^\beta(x_i^\alpha, P_L)$.

For large amounts of phases $\alpha$ and $\beta$, the equality $\mu_A^\beta(T, P^\beta, x_A^\beta) = \mu_A^\alpha(T, P^\beta, x_A^\alpha)$ for regular solutions is

$$\overline{\mu}_A^\beta(T, P^\beta) + kT \ln x_A^\beta + \Omega^\beta(1-x_A^\beta)^2 = \overline{\mu}_A^\alpha(T, P^\beta) + kT \ln x_A^\alpha + \Omega^\alpha(1-x_A^\alpha)^2, \quad \Omega^{\beta(\alpha)} \equiv kT\omega^{\beta(\alpha)} \tag{32}$$

According to the definition of $x_{A,\infty}^\beta$,

$$\overline{\mu}_A^\beta(T, P^\beta) + kT \ln x_{A,\infty}^\beta + \Omega^\beta(1-x_{A,\infty}^\beta)^2 = \overline{\mu}_A^\alpha(T, P^\beta) \tag{33}$$

Combining Eqs. (32) and (33), we get

$$\ln \frac{x_A^\beta}{x_{A,\infty}^\beta} + \omega^\beta\!\left[(1-x_A^\beta)^2 - (1-x_{A,\infty}^\beta)^2\right] = \ln x_A^\alpha + \omega^\alpha(1-x_A^\alpha)^2 \tag{34}$$

which is the first Eq. (31) for the planar interface ($P_L = 0$), as it must.

If we assume $\overline{\mu}_A^\beta(T, P^\beta) = \overline{\mu}_A^\alpha(T, P^\beta)$ instead of Eq. (33) and put $\omega^\alpha = \omega^\beta \equiv \omega$, then we have

$$\ln x_A^\beta + \omega(1-x_A^\beta)^2 = \ln x_A^\alpha + \omega(1-x_A^\alpha)^2 \tag{35}$$

instead of Eq. (34); the same equations hold for component B. Hence, it is seen that the conditions $\overline{\mu}_i^\beta = \overline{\mu}_i^\alpha$ are equivalent to $x_{A,\infty}^\beta = x_{B,\infty}^\beta = 1$ and $C_A = C_B = 1$. These conditions together with $\omega^\alpha = \omega^\beta$ are fulfilled for transformations within the same state of aggregation – the separation of liquid or solid solutions with the same structure of both the phases, where the phases differ from each other only by composition; just these conditions were employed by Prigogine and Defay [29] in the analytical description of this phenomenon. In the model of regular solution, the binodal and spinodal were calculated in Ref. [29] and thereby the metastable region between these curves was shown, where nucleation occurs. On the other hand, generally $\overline{\mu}_i^\beta \neq \overline{\mu}_i^\alpha$ for transformations between different state of aggregation, e.g. the crystallization of a melt (the equality holds only at the melting temperature of pure component $i$). So, the form of Eq. (31) is the most general.

The system of equations (30) determines the radius $R_*(x^\beta)$ and composition $x_*^\alpha(x^\beta)$ of a critical nucleus for the given composition of the mother phase. For deriving the dependence $x_*^\alpha(x^\beta)$, we denote $x_A \equiv x$, $x_B = 1 - x$ and neglect the dependence of $f_i^\alpha$ on $P^\alpha$; then divide the first Eq. (30) by the second one and denote $\gamma \equiv \upsilon_A^\alpha / \upsilon_B^\alpha$. After simple transformations, one obtains

$$\frac{x^\alpha}{(1-x^\alpha)^\gamma}\frac{f_A^\alpha(x^\alpha)}{\left[f_B^\alpha(x^\alpha)\right]^\gamma} = C\,\frac{x^\beta}{(1-x^\beta)^\gamma}\frac{f_A^\beta(x^\beta)}{\left[f_B^\beta(x^\beta)\right]^\gamma}\,, \quad C(T) \equiv \frac{\left[f_{B,\infty}^\beta x_{B,\infty}^\beta\right]^\gamma}{f_{A,\infty}^\beta x_{A,\infty}^\beta} \tag{36}$$

This equation implicitly gives the desired dependence $x_*^\alpha(x^\beta)$. The critical radius then can be found from any of Eqs. (30), say, the first:

$$R_*(x^\beta) = \frac{2\sigma\upsilon_A^\alpha}{kT}\left\{\ln\frac{f_A^\beta(x^\beta)x^\beta}{f_{A,\infty}^\beta x_{A,\infty}^\beta} - \ln f_A^\alpha(x^\alpha)x_*^\alpha(x^\beta)\right\}^{-1} \tag{37}$$

For a regular solution, Eq. (36) acquires the following form:

$$\frac{x^\alpha}{(1-x^\alpha)^\gamma}\exp\!\left\{\omega^\alpha\!\left[(1-x^\alpha)^2 - \gamma(x^\alpha)^2\right]\right\} = C_{reg}\,\frac{x^\beta}{(1-x^\beta)^\gamma}\exp\!\left\{\omega^\beta\!\left[(1-x^\beta)^2 - \gamma(x^\beta)^2\right]\right\},$$

$$C_{reg}(T) \equiv \frac{(x_{B,\infty}^\beta)^\gamma}{x_{A,\infty}^\beta}\exp\!\left\{\omega^\beta\!\left[\gamma(1-x_{B,\infty}^\beta)^2 - (1-x_{A,\infty}^\beta)^2\right]\right\} \tag{38}$$

## 2.4. Mother phase dilute with respect to A, nucleus dilute with respect to B

Differently form the previous case, the composition cannot be arbitrary here, so that the integration of Eq. (28) differs by choosing the initial state for component B. We have for the chemical potentials of both components in both the phases:

$$\mu_A^\beta = \psi^\beta(P^\beta, T) + kT \ln x_A^\beta\,, \quad \mu_A^\alpha = \overline{\mu}_A^\alpha(P^\alpha, T) + kT \ln x_A^\alpha$$

$$\mu_B^\beta = \overline{\mu}_B^\beta(P^\beta, T) + kT \ln x_B^\beta\,, \quad \mu_B^\alpha = \psi^\alpha(P^\alpha, T) + kT \ln x_B^\alpha \tag{39}$$

where by condition $x_A^\beta$, $x_B^\alpha \ll 1$ and $x_A^\alpha$, $x_B^\beta \sim 1$; $\psi^{\alpha(\beta)}$ is not a pure-component chemical potential.

Eq. (28) for component A is integrated in the same way, as above: from the initial state ($P_L = 0$, $x_A^\alpha = 1$, $x_A^\beta = x_{A,\infty}^\beta$) to the current state ($P_L$, $x_A^\alpha$, $x_A^\beta$); the quantity $x_{A,\infty}^\beta(T)$ has the same meaning, as before. As a result, we obtain Eq. (29a) which has the following form, in view of Eq. (39):

$$\ln\frac{x_A^\beta}{x_{A,\infty}^\beta} - \ln x_A^\alpha = \frac{\upsilon^\alpha P_L}{kT} \tag{40a}$$

For component B, we change the order of $\alpha$ and $\beta$ in Eq. (28), $\upsilon_i^\alpha dP_L - \dot{\mu}_i^\alpha dx_i^\beta + \dot{\mu}_i^\alpha dx_i^\alpha = 0$, and choose the initial state ($P_L = 0$, $x_B^\beta = 1$, $x_B^\alpha = x_{B,\infty}^\alpha$), where $x_{B,\infty}^\alpha(T)$ is the fraction of component B in phase $\alpha$ over the planar interface of pure B in phase $\beta$. The path of integration is as follows: (i) $P_L$ changes from 0 to the current $P_L$ at $x_B^\beta = 1$ and $x_B^\alpha = x_{B,\infty}^\alpha$; (ii) $x_B^\beta$ changes from 1 to the current $x_B^\beta$ at the given $P_L$ and $x_B^\alpha = x_{B,\infty}^\alpha$; (iii) $x_B^\alpha$ changes from $x_{B,\infty}^\alpha$ to the current $x_B^\alpha$ at the given $P_L$ and $x_B^\beta$. As a result, one obtains

$$\tilde{\upsilon}_B^\alpha P_L - \left[\mu_B^\beta(x_B^\beta) - \overline{\mu}_B^\beta\right] + \left[\mu_B^\alpha(x_B^\alpha) - \mu_B^\alpha(x_{B,\infty}^\alpha)\right] = 0 \qquad (40b)$$

where $\tilde{\upsilon}_B^\alpha$ is the partial volume of component B in phase $\alpha$ upon dilution; in view of our basic assumption of composition independence of $\upsilon_i$, it is equal to the specific volume $\upsilon_B^\alpha$. With account for Eq. (39), Eq. (40b) is as follows:

$$\ln\frac{x_B^\alpha}{x_{B,\infty}^\alpha} - \ln x_B^\beta = \frac{\upsilon_B^\alpha P_L}{kT} \qquad (40c)$$

Finally,

$$\begin{cases} \dfrac{x_A^\beta}{x_{A,\infty}^\beta} = x_A^\alpha \exp\left(\dfrac{\upsilon_A^\alpha P_L}{kT}\right) \\ x_B^\beta = \dfrac{x_B^\alpha}{x_{B,\infty}^\alpha}\exp\left(\dfrac{\upsilon_A^\alpha P_L}{kT}\right) \end{cases}, \quad \begin{cases} \dfrac{x^\beta}{x_{A,\infty}^\beta} = x^\alpha \exp\left(\dfrac{\upsilon_A^\alpha P_L}{kT}\right) \\ 1 - x^\beta = \dfrac{1 - x^\alpha}{x_{B,\infty}^\alpha}\exp\left(\dfrac{\upsilon_A^\alpha P_L}{kT}\right) \end{cases} \qquad (41)$$

where $x_A \equiv x$ and $x_B = 1 - x$ was put.

For the planar interface ($P_L = 0$),

$$\begin{cases} x^\beta / x_{A,\infty}^\beta = x^\alpha \\ 1 - x^\beta = (1 - x^\alpha) / x_{B,\infty}^\alpha \end{cases} \qquad (42)$$

These equations can be derived directly from Eq. (39) with the use of $x_{A,\infty}^\beta$ and $x_{B,\infty}^\alpha$ definition, as it was done above for a regular solution. From Eq. (42), the compositions of two coexisting bulk dilute solutions are determined:

$$x^\alpha = \frac{1 - x_{B,\infty}^\alpha}{1 - x_{A,\infty}^\beta x_{B,\infty}^\alpha} \approx 1 - x_{B,\infty}^\alpha(1 - x_{A,\infty}^\beta), \quad x^\beta = x_{A,\infty}^\beta(1 - x_{B,\infty}^\alpha), \quad x_B^\alpha = x_{B,\infty}^\alpha(1 - x_{A,\infty}^\beta) \qquad (43)$$

up to quadratic terms.

In order to obtain the composition $x_*^\alpha(x^\beta)$ of critical nucleus, we take the logarithm of Eq. (41), then divide the first equation by the second one and denote $\gamma \equiv \upsilon_A^\alpha / \upsilon_B^\alpha$. After transformations, one obtains

$$\frac{x^\alpha}{(1 - x^\alpha)^\gamma} = C\,\frac{x^\beta}{(1 - x^\beta)^\gamma}, \quad C(T) \equiv \frac{1}{x_{A,\infty}^\beta (x_{B,\infty}^\alpha)^\gamma} \qquad (44)$$

The critical radius is then obtained from Eq. (41) with the use of $x_*^\alpha(x^\beta)$, as before.

The critical radius also can be obtained directly form the system of equations (41). We express $x^\alpha$ from the first equation and substitute it to the second one; after transformations, one obtains

$$x^\beta = \frac{x_{A,\infty}^\beta \exp(\upsilon_A^\alpha P_L / kT)\left[1 - x_{B,\infty}^\alpha \exp(-\upsilon_B^\alpha P_L / kT)\right]}{1 - x_{A,\infty}^\beta x_{B,\infty}^\alpha \exp(\Delta\upsilon^\alpha P_L / kT)}, \quad \Delta\upsilon^\alpha \equiv \upsilon_A^\alpha - \upsilon_B^\alpha \qquad (45a)$$

Up to quadratic terms,

$$x^{\beta} = x_{A,\infty}^{\beta} \exp(\upsilon_A^{\alpha} P_L / kT)\left[1 - x_{B,\infty}^{\alpha} \exp(-\upsilon_B^{\alpha} P_L / kT)\right] \qquad (45b)$$

from where $R_*(x^{\beta})$ is determined. If component B is not soluble in A, then $x_{B,\infty}^{\alpha} = 0$ and Eqs. (45a, b) convert to Eq. (15) for the precipitation of pure component A.

Equations for the above quantities $x_{\infty}(T)$, $x_{A,\infty}^{\beta}(T)$, $x_{B,\infty}^{\beta}(T)$, etc. are found from the equality of chemical potentials of the corresponding bulk phases. E.g., for $x_{A,\infty}^{\beta}(T)$ in Eq. (29a), we have by definition $\mu_A^{\beta}(T, P^{\beta}, x_{A,\infty}^{\beta}) = \overline{\mu}_A^{\alpha}(T, P^{\beta})$. The differential form of this equation, $d\mu_A^{\beta}(T, P^{\beta}, x_{A,\infty}^{\beta}) = d\overline{\mu}_A^{\alpha}(T, P^{\beta})$ at constant pressure $P^{\beta}$ reads

$$-s_A^{\beta} dT + \dot{\mu}_A^{\beta} dx_{A,\infty}^{\beta} = -s_A^{\alpha} dT, \quad \dot{\mu}_A^{\beta} dx_{A,\infty}^{\beta} = \frac{q_A^{(\alpha\beta)}(T)}{T} dT, \quad q_A^{(\alpha\beta)}(T) = T(s_A^{\beta} - s_A^{\alpha}) \qquad (46)$$

where $q_A^{(\alpha\beta)}$ is the heat of transition "$\beta$ (solution) $\rightarrow \alpha$ (pure A)" for A atom. From this equation, the dependence $x_{A,\infty}^{\beta}(T)$ is determined. The examples of similar equations are given in Refs. [29, 30].

## 2.5. The work of nucleus formation

The work of a near-critical nucleus formation is given by Eq. (1). The work $W_*$ of critical nucleus formation is given by the familiar Gibbs equation

$$W_* = \frac{1}{3}\sigma A_* = \frac{4\pi}{3}\sigma \mathcal{R}_*^2 \qquad (47)$$

where the critical radius is given by the above equations.

The second differential of the work (the second summand in Eq. (1)) was calculated in Ref. [1] for a binary droplet; this calculation is also valid for the present theory. The variables ($V$, $x$) – the nucleus volume and composition $x \equiv x_A^{\alpha}$ - are used here as the variables of nucleus description in the two-dimensional problem. So, the matrix $\mathbf{H}$ (the coefficients $h_{ik}$) is

$$\mathbf{H} = \begin{pmatrix} -\dfrac{P_L^*}{3V_*} & 0 \\ 0 & N_* \dfrac{\dot{\mu}_{A,*}^{\alpha}}{1 - x_*} \end{pmatrix}, \quad \dot{\mu}_A^{\alpha} \equiv \left(\frac{\partial \mu_A^{\alpha}}{\partial x}\right)_{T, P^{\alpha}} \qquad (48)$$

In the CNT approximation (a homogeneous nucleus with bulk properties), the matrix $\mathbf{H}$ is diagonal, when the composition $x$ is used as a variable of nucleus description; it is non-diagonal, if the variables ($N_1$, $N_2$) are employed [1].

It was mentioned above that the composition of a two-component precipitate can fluctuate around the critical value $x_*$; the element $h_{xx}$ of Eq. (48) just determines the variance of this fluctuation: $\left\langle (x - x_*)^2 \right\rangle = kTh_{xx}^{-1}$. The positiveness of the element $h_{xx}$ is ensured by the thermodynamic condition of stability $\dot\mu_A^\alpha > 0$. The thermodynamic limit $h_{xx} \to \infty$ corresponds to transition to the single component

$(V)$-theory; accordingly, it is realized for pure A ($x_* \to 1$) or B ($x_* \to 0$) component. In the latter case,

we have for a dilute solution $\mu_A^\alpha = \psi(T, P^\alpha) + kT \ln x$, $\dot\mu_{A,*} = kT / x_* \to \infty$ at $x_* \to 0$.

Transformations in solid state (SS) can create elastic stresses which affect the nucleation kinetics. The work of nucleus formation in Ref. [1] and Eq. (48) resulting from it do not take into account this effect and therefore they can be applied only to the cases, where it is not essential. The detailed analysis of this phenomenon and the overview of works on this topic are given by Christian [31].

## 3. Equations of nucleus growth

### 3.1. The growth of a precipitate from ideal or dilute solution

At first, we consider the growth of a one-component (A) precipitate from binary solution. The concentration $c_A = cx_A$ - the number of A atoms in unit volume - is employed in this Section; $c$ is the total number of atoms in unit volume. The flux of A atoms to the nucleus of radius $R$ across the interface is $j_+ = 4\pi R^2 \nu_A l c_R^A$, where $\nu_A$ is the probability of jump across the interface per unit time, $l$ is the mean length of jump, and $c_R^A$ is the concentration of A atoms near the interface; the similar quantity enters the Russell model [24] mentioned above, however, for other purpose. The reverse flux $j_-$ is assumed to be the same, as in the *equilibrium* of the nucleus of radius $R$ with the mother phase:

$j_- = j_-^e = j_+^e = 4\pi R^2 l \nu_A c_e^A(R)$, where $c_e^A(R)$ is the equilibrium concentration of component A for the nucleus of radius $R$; it is given by Eqs. (10), (24) and their particular cases derived above. So, the net flux is

$$\dot N_A \equiv j = j_+ - j_- = 4\pi R^2 l \nu_A (c_R^A - c_e^A) \tag{49}$$

On the other hand, the flux $j$ can be found from the solution of stationary diffusion problem

$\Delta c_A(r) = 0$ with boundary conditions $c_A(\infty) = c_0^A$ and $c_A(R) = c_R^A$, where $c_0^A$ is the given concentration of binary solution: $c_A(r) = c_0^A - (c_0^A - c_R^A)R / r$ and

$$j = 4\pi R D_0^A (c_0^A - c_R^A) \tag{50}$$

where $D_0^A$ is the diffusion coefficient of component A in the solution. Comparing this equation to Eq. (49), we find the quantity $c_R^A$:

$$c_R^A = \frac{D_0^A c_0^A + l \nu_A R c_e^A}{D_0^A + l \nu_A R} \tag{51}$$

For the critical nucleus, $R = R_*$, we have by definition $c_e^A = c_0^A$ and hence $c_R^A = c_0^A$, as it must; according to Eq. (15), $c_e^A$ decreases with increasing $R$, thus $c_e^A < c_0^A$ for $R > R_*$.

It should be noted that the boundary condition $c_A(R) = c_e^A$ is often employed in literature [25]. However, there is a contradiction between kinetics and thermodynamics in this point: on the one hand, there is the flux of A atoms towards the nucleus, in view of the above inequality $c_e^A < c_0^A$; on the other hand, the nucleus cannot grow under this condition according to thermodynamics – it is in equilibrium with the mother phase. For this reason, the quantity $c_R^A$ is used here which is self-consistently determined. Substituting Eq. (51) in Eq. (50), we get finally

$$\dot{N}_A = 4\pi R D_0^A \Gamma_A (c_0^A - c_e^A), \quad \Gamma_A \equiv \frac{\gamma_A}{1+\gamma_A}, \quad \gamma_A \equiv \frac{l\nu_A R}{D_0^A} \tag{52a}$$

Obviously, the same analysis can be applied to B atoms in considering the growth of a two-component precipitate, and

$$\dot{N}_B = 4\pi R D_0^B \Gamma_A (c_0^B - c_e^B), \quad \Gamma_B \equiv \frac{\gamma_B}{1+\gamma_B}, \quad \gamma_B \equiv \frac{l\nu_B R}{D_0^B} \tag{52b}$$

where $c_e^A(R,x)$ and $c_e^B(R,x)$ are given by Eqs. (30), (31), and (41). It should be emphasized that $c_e^A(R,x) + c_e^B(R,x) \neq 1$ for $R \neq R_*$; the equality holds only for $R = R_*$, when $c_e^A(R_*, x_*) = c_0^A$ and $c_e^B(R_*, x_*) = c_0^B$. Eqs. (52a, b) determine the kinetics of evolution of a two-component precipitate – the change in its size and composition.

### 3.2. The growth of a precipitate from non-ideal solution

According to the thermodynamic theory of diffusion based on linear non-equilibrium thermodynamics [31, 32], the driving force of a diffusion process is the gradient of chemical potential, rather than the concentration gradient. As a consequence, the dependence of the diffusion coefficient on concentration arises in a non-ideal solution [32, 33]:

$$D(x) = D_0 \left[ 1 + \frac{\partial \ln f(x)}{\partial \ln x} \right] \tag{53}$$

where $f(x)$ is the activity employed above; the index A is omitted for brevity. For a regular solution, this gives

$$D(x) = D_0 \left[ 1 - 2\omega x(1-x) \right] \tag{54}$$

Stationary diffusion equation is

$$\frac{1}{r^2}\frac{d}{dr}\left[ r^2 D(x) \frac{dx}{dr} \right] = 0, \quad D(x)\frac{dx}{dr} = \frac{k_1}{r^2} \tag{55a}$$

Integration of this equation, in view of Eq. (54), gives

$$D_0 \left[ x - \omega x^2 + \frac{2}{3}\omega x^3 \right] = -\frac{k_1}{r} + k_2 \tag{55b}$$

The above boundary conditions $x(\infty) = x_0$ and $x(R) = x_R$ determine the integration constants:

$$k_2 = D_0 \left[ x_0 - \omega x_0^2 + \frac{2}{3} \omega x_0^3 \right], \quad k_1 = RD_0 \left[ (x_0 - x_R) - \omega (x_0^2 - x_R^2) + \frac{2}{3} \omega (x_0^3 - x_R^3) \right] \tag{55c}$$

The desired flux of A atoms towards the nucleus, in view of Eq. (55a), is $j = 4\pi R^2 c k_1 / R^2 = 4\pi c k_1$,

$$j = 4\pi RD_0 c \left[ (x_0 - x_R) - \omega (x_0^2 - x_R^2) + \frac{2}{3} \omega (x_0^3 - x_R^3) \right] \tag{56}$$

Returning to concentrations and comparing this equation to Eq. (49), we have

$$\gamma (c_R - c_e) = (c_0 - c_R) - \frac{\omega}{c} (c_0^2 - c_R^2) + \frac{2}{3} \frac{\omega}{c^2} (c_0^3 - c_R^3) \tag{57a}$$

Denoting $c_R - c_e \equiv z$ and $c_0 - c_e \equiv y$, we see that this equation implicitly determines the function $z(y)$, if we note that $c_0 - c_R = (c_0 - c_e) - (c_R - c_e) = y - z$, $c_0^2 - c_R^2 = y^2 + 2c_e y - z^2 - 2c_e z$, etc. The full representation of Eq. (57a) after rearrangements is as follows:

$$\left[ (\gamma + 1) - \frac{2\omega c_e}{c} + \frac{2\omega c_e^2}{c^2} \right] z + \left[ \frac{2\omega c_e}{c^2} - \frac{\omega}{c} \right] z^2 + \frac{2}{3} \frac{\omega}{c^2} z^3 = \left[ 1 - \frac{2\omega c_e}{c} + \frac{2\omega c_e^2}{c^2} \right] y + \left[ \frac{2\omega c_e}{c^2} - \frac{\omega}{c} \right] y^2 + \frac{2}{3} \frac{\omega}{c^2} y^3$$

$$\tag{57b}$$

It will be seen later that only the first (linear) term of the expansion $z(y) = z'(0) y$ is sufficient for our purpose; $y = 0$ corresponds to the critical nucleus, $c_e = c_0 = c_R$, hence, $z(0) = 0$. The derivative $z'(y) = dz / dy$ can be easily found by differentiating both sides of Eq. (57b) with respect to $y$. Its value at zero (with the restored index A) is

$$z'(0) = \frac{\widetilde{\gamma}_A}{\gamma_A + \widetilde{\gamma}_A}, \quad \widetilde{\gamma}_A \equiv 1 - \frac{2\omega c_e^A}{c} \left( 1 - \frac{c_e^A}{c} \right) \tag{57c}$$

Substituting $z(y) = z'(0) y$ for $(c_R^A - c_e^A)$ in Eq. (49), we get finally after simple transformations

$$\dot{N}_A = 4\pi RD_0^A \widetilde{\Gamma}_A (c_0^A - c_e^A), \quad \widetilde{\Gamma}_A \equiv \frac{\gamma_A \widetilde{\gamma}_A}{\gamma_A + \widetilde{\gamma}_A} \tag{58a}$$

Eq. (53) has a symmetric form for both components: $D_B(x) = D_0^B \left[ 1 + \partial \ln f_B(x) / \partial \ln x_B \right]$, so that the above analysis can be applied to component B without any changes, and

$$\dot{N}_B = 4\pi RD_0^B \widetilde{\Gamma}_B (c_0^B - c_e^B), \quad \widetilde{\Gamma}_B \equiv \frac{\gamma_B \widetilde{\gamma}_B}{\gamma_B + \widetilde{\gamma}_B}, \quad \widetilde{\gamma}_B \equiv 1 - \frac{2\omega c_e^B}{c} \left( 1 - \frac{c_e^B}{c} \right) \tag{58b}$$

Eqs. (58a, b) determine the kinetics of growth of a two-component precipitate from a regular solution. For an ideal solution, $\omega = 0$, $\widetilde{\gamma}_A = \widetilde{\gamma}_B = 1$ and $\widetilde{\Gamma}_A = \Gamma_A$, $\widetilde{\Gamma}_B = \Gamma_B$.

In conclusion, the analysis of coefficients $\gamma_i$ and $\Gamma_i$ should be done. Representing the diffusion coefficient as $D_0^i = l^2 \nu_i^{(bulk)}$, where $\nu_i^{(bulk)}$ is the frequency of $i$ atom jumps in the bulk mother phase (it includes all necessary quantities such as the coordination number, the correlation factor, etc. [31]), we get

$$\gamma_i = \frac{R}{l} \frac{\nu_i}{\nu_i^{(bulk)}} \tag{59}$$

The two limiting cases are as follows. (i) The interface-controlled growth with respect to component $i$: $\gamma_i << 1$, $\Gamma_i = \gamma_i$, and

$$\dot{N}_i = 4\pi R^2 l \nu_i (c_0^i - c_e^i) \tag{60a}$$

This case corresponds to fast diffusion in the bulk and slow kinetics at the interface; the latter determines the growth with respect to component $i$. (ii) The diffusion-limited growth with respect to component $i$: $\gamma_i >> 1$, $\Gamma_i = 1$, and

$$\dot{N}_i = 4\pi R D_0^i (c_0^A - c_e^A) \tag{60b}$$

This case corresponds to fast kinetics at the interface and slow diffusion in the bulk; the slowest process determines the growth, as before.

Just Eq. (60b) is usually employed in literature [25] for describing the growth from solution; the interfacial kinetics falls out from consideration. As is seen from above, the appearance of quantities $\gamma_i$ in the present theory is due to introducing the quantity $c_R^i$ and employing the boundary condition $c_A(R) = c_R^A$ instead of $c_A(R) = c_e^A$; just $\gamma_i$ allow revealing these limiting cases. In view of natural condition $R > R_a \approx l/2$, where $R_a$ is the atomic radius, the interface-controlled growth requires $\nu_i << \nu_i^{(bulk)}$. For a sufficiently small (nanosized) nucleus, $R$ is few times $l/2$, so that the diffusion-limited growth requires the inverse condition $\nu_i >> \nu_i^{(bulk)}$. Likely, some intermediate case occurs for an actual growth.

The similar analysis holds for the quantities with tilde. (i) The interface-controlled growth: $\gamma_i << \tilde{\gamma}_i$, $\tilde{\Gamma}_i = \gamma_i$, and Eq. (58a) goes to Eq. (60a). (ii) The diffusion-limited growth: $\gamma_i >> \tilde{\gamma}_i$, $\tilde{\Gamma}_i = \tilde{\gamma}_i$, and

$$\dot{N}_i = 4\pi R D_0^i \tilde{\gamma}_i (c_0^i - c_e^i) \tag{60c}$$

As $\tilde{\gamma}_i < 1$, the condition for case (i) is stronger, whereas the condition for case (ii) is weaker, than in the previous consideration.

Finally, it should be noted that always $\tilde{\gamma}_i > 0$, though it involves a negative term. The equation for spinodal curve $T(x)$ of a regular solution is [29] $kT/2\Omega = x(1-x)$, or $2\omega x(1-x) = 1$; it is the same for $x_A$ and $x_B$. This equation for $\omega > 2$ (which corresponds to $T < T_c = \Omega/2k$, $T_c$ is the critical temperature) yields two spinodal values $x_1^s = (1 - \sqrt{1 - 2/\omega})/2$ and $x_2^s = (1 + \sqrt{1 - 2/\omega})/2$. The composition of mother phase can be $x_0 < x_1^s$ or $x_0 > x_2^s$; the region $x_1^s < x_0 < x_2^s$ is unstable. Obviously, the equilibrium compositions $x_e^i$ are located in the allowed regions, $x_e < x_1^s$ and $x_e > x_2^s$, so that $\tilde{\gamma}_i = 1 - 2\omega x_e^i (1 - x_e^i) > 0$.

### 3.3. Nucleus motion equations in the $(V, x)$-space

Eq. (2) in the two-dimensional $(V, x)$-space has the following form:

$$\begin{cases} \dot{V} = -z_{VV}(V - V_*) - z_{Vx}(x - x_*) \\ \dot{x} = -z_{xV}(V - V_*) - z_{xx}(x - x_*) \end{cases} \quad (61)$$

The equation for $\dot{x}$ can be also represented as

$$\dot{x} = a_x \dot{V} - \lambda_{xx}(x - x_*) \quad (62)$$

From comparison, the matrix $\mathbf{Z}$ is

$$\mathbf{Z} = \begin{pmatrix} z_{VV} & z_{Vx} \\ a_x z_{VV} & a_x z_{Vx} + \lambda_{xx} \end{pmatrix} \quad (63)$$

The condition of symmetry of the matrix $\mathbf{B} = kT\,\mathbf{ZH}^{-1}$ (Onsager's reciprocal relation) results in equation

$$z_{Vx} h_{xx}^{-1} = a_x z_{VV} h_{VV}^{-1}$$

from where

$$a_x = \frac{z_{Vx}}{z_{VV}} \frac{h_{VV}}{h_{xx}} \quad (64)$$

In the one-dimensional ($V$)-theory, the equation of motion is

$$\dot{V} = -z_{VV}(V - V_*) \quad (65)$$

## 4. Results and discussion

### 4.1. Nucleation rate of a one-component precipitate

Calculations can be done for the general case of Eq. (10), however, they are given below by the example of explicit Eq. (13). According to Eq. (58a),

$$\dot{V} = \upsilon_A^\alpha \dot{N}_A = 4\pi R \upsilon_A^\alpha D_0^A \tilde{\Gamma}_A c(x_0^A - x_e^A) \quad (66)$$

where $x_e^A(R)$ is given by Eq. (13),

$$x_e^A \exp[\omega^\beta(1 - x_e^A)^2] = C \exp\left(\frac{\upsilon_A^\alpha P_L}{kT}\right) \quad (67)$$

whereas $x_0^A$ is the current fraction of component A in the solution.

All we need for calculating the nucleation rate is the quantity $z_{VV}$ in Eq. (65), $z_{VV} = -(d\dot{V}/dV)_*$. As is seen from Eq. (66), this derivative is reduced to calculating the derivative $(dx_e^A/dV)_*$ ($x_e^A = x_0^A$ at $V = V_*$, so that the derivative of the multiplier in front of brackets does not contribute to $z_{VV}$). The following useful equalities are employed below:

$$\left(\frac{d^2\sigma A}{dV^2}\right)_* = \left(\frac{dP_L}{dV}\right)_* = \left(\frac{dP_L}{dR}\frac{dR}{dV}\right)_* = -\frac{P_L^*}{3V_*} = h_{VV} \quad (68a)$$

according to Eq. (48), and therefore

$$\left(\frac{\partial x_e^A}{\partial V}\right)_* = \left(\frac{\partial x_e^A}{\partial P_L}\frac{\partial P_L}{\partial V}\right)_* = \left(\frac{\partial x_e^A}{\partial P_L}\right)_* h_{VV} \tag{68b}$$

To get from Eq. (67) the derivative $dx_e^A / dP_L$, we differentiate both sides of this equation with respect to $P_L$:

$$\frac{dx_e^A}{dP_L} = C\frac{\upsilon_A^\alpha}{kT}e^{\frac{\upsilon_A^\alpha P_L}{kT}}\frac{e^{-\omega^\beta(1-x_e^A)^2}}{1-2\omega^\beta x_e^A(1-x_e^A)} \tag{69a}$$

Taking this derivative at $V = V_*$, we substitute $x_e^A = x_0^A$ and employ Eq. (67):

$$\left(\frac{dx_e^A}{dP_L}\right)_* = \frac{\upsilon_A^\alpha}{kT}\frac{x_0^A}{1-2\omega^\beta x_0^A(1-x_0^A)} = \frac{\upsilon_A^\alpha}{kT}\frac{x_0^A}{\widetilde{\gamma}_A^*} \tag{69b}$$

In view of the equality

$$\frac{\widetilde{\Gamma}_A^*}{\widetilde{\gamma}_A^*} = \frac{\gamma_A^*}{\gamma_A^* + \widetilde{\gamma}_A^*} \equiv \Gamma_A' \tag{69c}$$

we get finally

$$z_{VV} = \frac{4\pi R_* D_0^A (\upsilon_A^\alpha)^2 \Gamma_A' c\, x_0^A}{kT}h_{VV} = \frac{1}{kT}b_{VV}h_{VV}, \quad b_{VV} = 4\pi R_* D_0^A(\upsilon_A^\alpha)^2 c\, x_0^A \tag{70}$$

where $b_{VV}$ is the diffusivity of the Fokker-Planck equation in the $(V)$-space. It should be recalled that $\Gamma_A'$ is a function of composition $x_0^A$, since $\widetilde{\gamma}_A^*$ is a function of $x_0^A$; $\Gamma_A' = \Gamma_A^* = \gamma_A^* / (1 + \gamma_A^*)$ for ideal solution, $\omega^\beta = 0$ and $\widetilde{\gamma}_A^* = 1$.

Substituting $h_{11} \equiv h_{VV}$ and $\kappa_1 = z_{VV}$ in Eq. (5), we find the steady state nucleation rate:

$$I = C_0\sqrt{\frac{|h_{VV}|}{2\pi kT}}b_{VV}e^{-\frac{W_*}{kT}} = C_0\sqrt{\frac{\sigma}{kT}}\frac{2D_0^A(\upsilon_A^\alpha)^2\Gamma_A' c\, x_0^A}{R_*}e^{-\frac{W_*}{kT}} \tag{71a}$$

which is the one-dimensional Zel'dovich-Frenkel equation with the diffusivity calculated within the present approach, Eq. (70). $C_0$ is the normalizing factor of the equilibrium distribution function $F_e(V) = F_e(N)dN / dV = F_e(N) / \upsilon_A^\alpha$. So, if the $(N)$-space is used, we put $C_0 = C_0^{(N)} / \upsilon_A^\alpha$ and $b_{VV} = (\upsilon_A^\alpha)^2 b_{NN}$, $h_{VV} = h_{NN} / (\upsilon_A^\alpha)^2$.

In the case of interfaced-controlled growth, $\gamma_A^* << \widetilde{\gamma}_A^*$, we have $\Gamma_A' = \gamma_A^* / \widetilde{\gamma}_A^*$,

$$b_{VV} = \frac{4\pi R_*^2 l\, \upsilon_A (\upsilon_A^\alpha)^2 c\, x_0^A}{\widetilde{\gamma}_A^*}, \qquad I = C_0\sqrt{\frac{\sigma}{kT}}\frac{2(\upsilon_A^\alpha)^2 l\, \upsilon_A c\, x_0^A}{\widetilde{\gamma}_A^*}e^{-\frac{W_*}{kT}} \tag{71b}$$

and the preexponential factor does not include $R_*$. The same is true for the droplet nucleation [20]: the Zeldovich factor is proportional to $R_*^{-2}$, whereas the diffusivity is proportional to $R_*^2$; both these factors cancel each other.

In the case of diffusion-limited growth $\gamma_A^* >> \widetilde{\gamma}_A^*$, $\Gamma_A' = 1$ in Eqs. (70), (71a) and the diffusivity is proportional to $R_*$; as a result, $R_*$ appears in the denominator. On the other hand, Eq. (71b) includes $\widetilde{\gamma}_A^*$ in the denominator (for a non-ideal solution). Formally, both $R_*$ and $\tilde{\gamma}_A^*$ vanish on the spinodal. However, the values of $R_*$ are bounded by $R_a$ from below; the formal thermodynamic understanding of spinodal as corresponding to $R_* = 0$ must be also corrected for this condition. At the same time, the nucleation rate sharply increases at high supersaturations (small $R_*$), which is the well known fact. So, Eqs. (71a, b) are physically correct.

## 4.2. Nucleation rate of a compound

It will be seen later that we should differ the roles of components here; let A and B be solute and solvent, respectively, so that the growth of compound is determined by component A:

$$\dot{V} = \frac{\upsilon_c^\alpha \dot{N}_A}{n} = \frac{4\pi R \upsilon_c^\alpha D_0^A \tilde{\Gamma}_A c}{n}(x_0^A - x_e^A) \qquad (72)$$

where the division by $n$ means that $n$ A atoms from their full flux contribute to the formation of one compound molecule of volume $\upsilon_C^\alpha = n\upsilon_A^\alpha + m\upsilon_B^\alpha$; it is implied that the required $m$ B atoms at once attach to the mentioned $n$ A atoms in this process. The quantity $x_e^A(R)$ is given by Eq. (26):

$$y(x_e^A) \equiv (x_e^A)^n (1 - x_e^A)^m \exp\left\{\omega^\beta \left[n(1 - x_e^A)^2 + m(x_e^A)^2\right]\right\} = C \exp\left(\frac{\upsilon_c^\alpha P_L}{kT}\right) \qquad (73)$$

Further algorithm is the same as above, Eqs. (65)-(71). Calculating the derivative $dx_e^A / dP_L$ and taking it at $V = V_*$, we substitute $x_e^A = x_0^A$ and employ Eq. (73); as a result,

$$\left(\frac{dx_e^A}{dP_L}\right)_* = \frac{\upsilon_c^\alpha}{kT} \frac{x_0^A(1 - x_0^A)}{\left[n(1 - x_0^A) - mx_0^A\right]\left[1 - 2\omega^\beta x_0^A(1 - x_0^A)\right]} = \frac{\upsilon_c^\alpha}{kT} \frac{x_0^A(1 - x_0^A)}{\left[n(1 - x_0^A) - mx_0^A\right]}\tilde{\gamma}_A^* \equiv \frac{\upsilon_c^\alpha}{kT}\varphi(x_0^A) \qquad (74a)$$

from where

$$z_{VV} = \frac{1}{kT}b_{VV}h_{VV} , \quad b_{VV} = \frac{4\pi R_* D_0^A (\upsilon_c^\alpha)^2 \Gamma_A' c}{n}\frac{x_0^A(1 - x_0^A)}{n(1 - x_0^A) - mx_0^A} \qquad (74b)$$

The condition of $dx_e^A / dP_L > 0$ and accordingly $b_{VV} > 0$ is $n(1 - x_0^A) - mx_0^A > 0$, from where

$$x_0^A < \frac{n}{n + m} \equiv x_c \qquad (75)$$

i.e. the fraction of component A in the solution must be less than in the compound; this inequality can be regarded as a criterion for component A to be solute. Eq. (73) for component B is obtained simply by substituting $x_e^A = 1 - x_e^B$; obviously, here the equality $x_e^A + x_e^B = 1$ holds, differently from the case of binary precipitate of variable composition. Hence, $(dx_e^B / dP_L)_* = -(dx_e^A / dP_L)_*$. This means that the flux of atoms B is opposite to the flux of atoms A, i.e. their excess near the compound takes place. The left-hand side of or Eq. (73) for components A and B is shown in Fig. 1. The curve is symmetric for $n = m$ and asymmetric for $n \neq m$. The maxima are located at spinodal values, if $x_1^s < x_c < x_2^s$; at $x_c$ and $x_2^s$, if $x_c < x_1^s$; at $x_1^s$ and $x_c$, if $x_c > x_2^s$. Eq. (74a) has singularities both on the spinodal and at $x_0^A = x_c$; the solution composition $x_0^A = x_c$ can be called degenerate. The derivative $(dx_e^A / dP_L)_*$ is positive on the ascending branch of the curve and negative on the descending one; the situation is mirror-symmetric for component B. In other words, when $(dx_e^A / dP_L)_* < 0$, component B as a solute determines the growth of compound.

From this reasoning and in view of the equality $\tilde{\gamma}_A^* = \tilde{\gamma}_B^* \equiv \tilde{\gamma}^*$, we get

$$I = \begin{cases} C_0 \sqrt{\dfrac{\sigma}{kT}} \dfrac{2D_0^A (\upsilon_c^\alpha)^2 \Gamma_A' c}{nR_*} \dfrac{x_0^A (1-x_0^A)}{n(1-x_0^A)-mx_0^A} e^{-\frac{W_*}{kT}}, & \text{A solute} \\[4mm] C_0 \sqrt{\dfrac{\sigma}{kT}} \dfrac{2D_0^B (\upsilon_c^\alpha)^2 \Gamma_B' c}{mR_*} \dfrac{x_0^B (1-x_0^B)}{m(1-x_0^B)-nx_0^B} e^{-\frac{W_*}{kT}}, & \text{B solute} \end{cases} \tag{76}$$

The regions where A or B determines the nucleation rate are shown in Fig. 1a'-c'. In the case $x_c < x_1^s$, there is an additional narrow region $x_c < x_0^A < x_1^s$, where B determines the nucleation (Fig. 1c'); the same holds for A in the case $x_c > x_2^s$. In the case of ideal solution, Fig. 1b", only the quantity $x_c$ separates regions A and B.

### 4.3. Nucleation rate of a binary precipitate

An equation for $\dot{V}$ in this case is

$$\dot{V} = \upsilon_A^\alpha \dot{N}_A + \upsilon_B^\alpha \dot{N}_B \tag{77a}$$

where $\dot{N}_i$ are given by Eqs. (58a, b):

$$\dot{V} = 4\pi Rc \left[ \upsilon_A^\alpha D_0^A \tilde{\Gamma}_A (x_0^A - x_e^A) + \upsilon_B^\alpha D_0^B \tilde{\Gamma}_B (x_0^B - x_e^B) \right] \tag{77b}$$

If the dependence of $\upsilon_i^\alpha$ on composition is taken into account, then the term $(\dot{\upsilon}_A^\alpha N_A + \dot{\upsilon}_B^\alpha N_B)$ must be added to the RHS of Eq. (77a). In view of the equalities $\dot{\upsilon}_A^\alpha = (\partial \upsilon_A^\alpha / \partial x)\dot{x}$ and $\dot{\upsilon}_B^\alpha = (\partial \upsilon_B^\alpha / \partial x)\dot{x}$, where $x \equiv x_A^\alpha$, we have

$$\dot{\upsilon}_A^\alpha N_A + \dot{\upsilon}_B^\alpha N_B = N \left[ \dot{\upsilon}_A^\alpha x + \dot{\upsilon}_B^\alpha (1-x) \right] = N\dot{x} \left[ x \frac{\partial \upsilon_A^\alpha}{\partial x} + (1-x) \frac{\partial \upsilon_B^\alpha}{\partial x} \right] = 0 \tag{77c}$$

since the expression in brackets is equal to zero, according to the familiar property of partial volumes [29]. So, the dependence of partial volumes on composition does not change basic Eq. (77a).

An equation for $\dot{x}$ is obtained by differentiating $x = N_A / N$ as follows:

$$\dot{x} = \frac{1-x}{N} \dot{N}_A - \frac{x}{N} \dot{N}_B \tag{78a}$$

$$\dot{x} = \frac{4\pi Rc}{N} \left[ (1-x) \upsilon_A^\alpha D_0^A \tilde{\Gamma}_A (x_0^A - x_e^A) - x \upsilon_B^\alpha D_0^B \tilde{\Gamma}_B (x_0^B - x_e^B) \right] \tag{78b}$$

The equations of nucleus motion now have the form of Eq. (61) and our aim is to find an explicit form of the matrix $\mathbf{Z}$, Eq. (63). Equations for $x_e^A(R,x)$ and $x_e^B(R,x)$ in Eqs. (77b) and (78b) are given by Eq. (31):

$$\begin{cases} x_e^A \exp\left[\omega^\beta (1-x_e^A)^2\right] = C_A x \exp\left[\dfrac{\upsilon_A^\alpha P_L}{kT} + \omega^\alpha (1-x)^2\right] \\ x_e^B \exp\left[\omega^\beta (1-x_e^B)^2\right] = C_B(1-x) \exp\left[\dfrac{\upsilon_B^\alpha P_L}{kT} + \omega^\alpha x^2\right] \end{cases} \tag{79}$$

Starting from calculating $z_{VV} = -(d\dot{V}/dV)_*$, we have to compute the derivatives $\partial x_e^A/\partial V$ and $\partial x_e^B/\partial V$. For this purpose, we use the known from analysis theorem on the differentiation of an implicit function. The function $x_e^A(P_L, x)$ is given implicitly by the relation $Y(x_e^A, P_L, x) = 0$, where

$$Y(x_e^A, P_L, x) = x_e^A \exp\left[\omega^\beta (1-x_e^A)^2\right] - C_A x \exp\left[\frac{\upsilon_A^\alpha P_L}{kT} + \omega^\alpha (1-x)^2\right] \tag{80a}$$

According to the mentioned theorem,

$$\frac{\partial x_e^A}{\partial P_L} = -\frac{\partial Y/\partial P_L}{\partial Y/\partial x_e^A}, \quad \frac{\partial x_e^A}{\partial x} = -\frac{\partial Y/\partial x}{\partial Y/\partial x_e^A} \tag{80b}$$

Calculating $\partial x_e^i/\partial P_L$, we then take it at the saddle point $(V_*, x_*)$, i.e. substitute $x_e^i = x_0^i$, as before, and employ Eq. (79); as a result,

$$\left(\frac{\partial x_e^i}{\partial P_L}\right)_* = \frac{\upsilon_i^\alpha}{kT} \frac{x_0^i}{1 - 2\omega^\beta x_0^i (1-x_0^i)}, \quad i = A, B \tag{81}$$

from where and in view of Eq. (68b)

$$z_{VV} = \frac{4\pi R_* c}{kT} \left[D_0^A (\upsilon_A^\alpha)^2 \Gamma_A' x_0 + D_0^B (\upsilon_B^\alpha)^2 \Gamma_B' (1-x_0)\right] h_{VV} \tag{82}$$

where $\Gamma_i' = \tilde{\Gamma}_i^*/\tilde{\gamma}^*$, $\tilde{\gamma}^* = 1 - 2\omega^\beta x_0^i (1-x_0^i)$, as before, and hereafter we put $x_0^A \equiv x_0$ and $x_0^B = 1-x_0$ in final equations. It is seen that Eq. (82) is a direct extension of Eq. (70) to binary case.

Further, we employ Eq. (77b) to calculate $z_{Vx} = -(d\dot{V}/dx)_*$; the derivatives $\partial x_e^A/\partial x$ and $\partial x_e^B/\partial x$ are computed according to the second Eq. (80b). Being taken at the saddle point with the use of Eq. (79), they have the following form:

$$\left(\frac{\partial x_e^A}{\partial x}\right)_* = \frac{1 - 2\omega^\alpha x_*(1-x_*)}{x_*} \frac{x_0^A}{1 - 2\omega^\beta x_0^A (1-x_0^A)} \tag{83a}$$

$$\left(\frac{\partial x_e^B}{\partial x}\right)_* = -\frac{1 - 2\omega^\alpha x_*(1-x_*)}{1-x_*} \frac{x_0^B}{1 - 2\omega^\beta x_0^B (1-x_0^B)} \tag{83b}$$

As a result,

$$z_{Vx} = 4\pi R_* c \left[1 - 2\omega^\alpha x_*(1-x_*)\right] \left\{ \frac{D_0^A \upsilon_A^\alpha \Gamma_A' x_0}{x_*} - \frac{D_0^B \upsilon_B^\alpha \Gamma_B' (1-x_0)}{1-x_*} \right\} \tag{84}$$

Eq. (78b) is used to obtain the elements of the second row. To get $z_{xV} = -(\partial \dot{x}/\partial V)_*$ the derivatives $\partial x_e^A/\partial V$ and $\partial x_e^B/\partial V$ found above are employed again:

$$z_{xV} = \frac{4\pi R_* c}{kTN_*}\left\{D_0^A v_A^\alpha \Gamma_A'(1-x_*)x_0 - D_0^B v_B^\alpha \Gamma_B' x_*(1-x_0)\right\}h_{VV}, \quad N_* = \frac{V_*}{v_A^\alpha x_* + v_B^\alpha(1-x_*)} \quad (85)$$

Similarly, the above derivatives $\partial x_e^A/\partial x$ and $\partial x_e^B/\partial x$ are employed to compute $z_{xx} = -(\partial \dot{x}/\partial x)_*$,

$$z_{xx} = \frac{4\pi R_* c\left[1 - 2\omega^\alpha x_*(1-x_*)\right]}{N_*}\left\{\frac{D_0^A \Gamma_A'(1-x_*)x_0}{x_*} + \frac{D_0^B \Gamma_B' x_*(1-x_0)}{1-x_*}\right\} \quad (86)$$

The element $z_{xV}$ was found above directly from basic Eq. (78b). On the other hand, its value is dictated by Eqs. (63) and (64): $z_{xV} = a_x z_{VV} = z_{Vx} h_{VV}/h_{xx}$, as a consequence of Onsager's principle. The element $h_{xx}$, Eq. (48), for a regular solution is

$$h_{xx} = N_* kT \frac{1 - 2\omega^\alpha x_*(1-x_*)}{x_*(1-x_*)} \quad (87)$$

With the use of this equation and Eq. (84), it is easy to see that the same Eq. (85) is obtained, i.e. the necessary condition of *self-consistency* of the theory is fulfilled - the thermodynamic and kinetic equations of the present approach are consistent with Onsager's principle.

With the use of notations

$$\xi_A \equiv \frac{4\pi R_* c\, v_A^\alpha \Gamma_A'}{N_*}x_0 D_0^A, \quad \xi_B \equiv \frac{4\pi R_* c\, v_B^\alpha \Gamma_B'}{N_*}(1-x_0)D_0^B \quad (88)$$

as well as Eq. (87), the matrix $\mathbf{Z}$ acquires the form

$$\mathbf{Z} = \frac{1}{kT}\begin{pmatrix} N_*\left[\xi_A v_A^\alpha + \xi_B v_B^\alpha\right]h_{VV} & \left[\xi_A(1-x_*) - \xi_B x_*\right]h_{xx} \\ \left[\xi_A(1-x_*) - \xi_B x_*\right]h_{VV} & \left[\dfrac{\xi_A}{v_A^\alpha}(1-x_*)^2 + \dfrac{\xi_B}{v_B^\alpha}x_*^2\right]\dfrac{h_{xx}}{N_*} \end{pmatrix} \quad (89)$$

It should be recalled that $R_*$ and $x_*$ are functions of $x_0$. Such a representation of $\mathbf{Z}$ is convenient to easy obtain the tensor of diffusivities of the Fokker-Planck equation in the $(V, x)$-space:

$$\mathbf{B} = kT\,\mathbf{Z}\mathbf{H}^{-1} = kT\begin{pmatrix} z_{VV}h_{VV}^{-1} & z_{Vx}h_{xx}^{-1} \\ z_{xV}h_{VV}^{-1} & z_{xx}h_{xx}^{-1} \end{pmatrix} \quad (90a)$$

$$\mathbf{B} = \begin{pmatrix} N_*\left[\xi_A v_A^\alpha + \xi_B v_B^\alpha\right] & \left[\xi_A(1-x_*) - \xi_B x_*\right] \\ \left[\xi_A(1-x_*) - \xi_B x_*\right] & \dfrac{1}{N_*}\left[\dfrac{\xi_A}{v_A^\alpha}(1-x_*)^2 + \dfrac{\xi_B}{v_B^\alpha}x_*^2\right] \end{pmatrix} \quad (90b)$$

It is symmetric, as it must; this fact again shows that the thermodynamic and kinetic equations used here result in the fulfillment of Onsager's principle. If we change the definition of $\xi_i$ as $\xi_i' = \xi_i N_* v_i^\alpha/3V_* kT$, the matrix $\mathbf{Z}$ elements expressed in terms of $\xi_i'$ are the same as for binary droplet nucleation in a mixture of two vapors, Ref. [1], Eq. (66) therein. This fact shows the universality of the present approach:

although the physics is different in these two systems (different equations of equilibrium and different equations of growth), the kinetic matrix $\mathbf{Z}$ has the same general form.

The remaining point in this part is to get the kinetic parameter $\lambda_{xx}$; according to Eq. (63), $\lambda_{xx} = z_{xx} - a_x z_{Vx}$ and $a_x = z_{xV} / z_{VV}$. From these equations, one obtains

$$\lambda_{xx} = \frac{h_{xx}}{N_* kT} \frac{(\upsilon^\alpha)^2}{\upsilon_A^\alpha \upsilon_A^\beta} \frac{\xi_A \xi_B}{\xi_A \upsilon_A^\alpha + \xi_B \upsilon_B^\alpha}, \quad \upsilon^\alpha = x_* \upsilon_A^\alpha + (1 - x_*) \upsilon_B^\alpha \tag{91}$$

The matrix $\mathbf{Z}$ for the nucleation of a precipitate dilute with respect to component B from mother phase dilute with respect to A is derived by the same algorithm with one simplification - $x_e^A(R, x)$ and $x_e^B(R, x)$ are explicit functions now, they are given by Eq. (41):

$$\begin{cases} x_e^A = C_A x \exp\left( \dfrac{\upsilon_A^\alpha P_L}{kT} \right) \\ x_e^B = C_B (1 - x) \exp\left( \dfrac{\upsilon_A^\alpha P_L}{kT} \right) \end{cases}, \quad C_A \equiv x_{A,\infty}^\beta, \; C_B \equiv x_{B,\infty}^\alpha \tag{92}$$

So, their derivatives with respect to $P_L$ and $x$ are calculated directly. Eqs. (52a, b) are substituted now in Eqs. (77a) and (78a); the element $h_{xx}$ has the form

$$h_{xx} = N_* \frac{kT}{x_*(1 - x_*)} \tag{93}$$

As a result of application of the above procedure, the same Eq. (89) is obtained for the matrix $\mathbf{Z}$, where $\xi_i$ are given by Eq. (88) with $\Gamma_i$, Eqs. (52a, b), instead of $\Gamma_i'$. In the approximation of diffusion-limited growth, $\Gamma_i = \Gamma_i' = 1$, so that the quantities $\xi_i$ become identical for both the problems; the matrices $\mathbf{Z}$ for the same $R_*$ differ by $x_*$ and $h_{xx}$. Since $x_*$ is close to unity in Eq. (93), this $h_{xx}$ is much greater, than $h_{xx}$ given by Eq. (87) for a regular solution.

The characteristic equation for the matrix $\mathbf{Z}$ is

$$\kappa^2 - (Sp\mathbf{Z})\kappa + \det \mathbf{Z} = 0, \quad Sp\mathbf{Z} = z_{VV} + z_{xx} = z_{VV} + \lambda_{xx} + a_x z_{Vx}; \quad \det \mathbf{Z} = z_{VV} \lambda_{xx} \tag{94}$$

The negative root is

$$\kappa_1 = \frac{1}{2} \left\{ Sp\mathbf{Z} - \sqrt{(Sp\mathbf{Z})^2 - 4\det \mathbf{Z}} \right\} \tag{95}$$

This quantity is substituted in Eq. (5) to get the steady state nucleation rate of a binary precipitate.

## 4.4. Kinetic limits

As is seen from the foregoing, the kinetics of binary nucleation in a condensed state within the CNT approximation is governed by the two parameters: $z_{VV}$ and $\lambda_{xx}$. The parameter $z_{VV}$ characterizes the rate of nucleus growth; in a one-dimensional problem, this is the relative change of the nucleus volume per unit time. The parameter $\lambda_{xx}$ characterizes *the rate of composition relaxation* at constant $V$; according to Eq. (62), $(\dot{x})_V = -\lambda_{xx}(x - x_*)$. Similarly to the thermodynamic limit $h_{xx} \to \infty$ discussed above, kinetic limits are also possible; they are determined by limiting relations between the kinetic parameters $z_{VV}$ and $\lambda_{xx}$.

The *unary nucleation limit* $\lambda_{xx} \to \infty$, or $\lambda_{xx} >> |z_{VV}|, a_x z_{Vx}$, leads to the conditions $(Sp\mathbf{Z})^2 >> |\det \mathbf{Z}|$ and $Sp\mathbf{Z} > 0$; Eq. (95) has the following asymptotics in this case:

$$\kappa_1 = \frac{\det \mathbf{Z}}{Sp\mathbf{Z}} = \frac{z_{VV}\lambda_{xx}}{z_{VV} + a_x z_{Vx} + \lambda_{xx}} = z_{VV} \tag{96}$$

which is the one-dimensional result. Indeed, a large value of $\lambda_{xx}$ means that any deviation of the nucleus composition from $x_*$ rapidly relaxes, i.e. the nucleus grows with *constant composition $x_*$*. In other words, the problem becomes one-dimensional; the variable $x$ falls out from consideration, only the variable $V$ remains. Alongside with the case of compound nucleation, the given case also can be attributed to Russel's model [24], although the physics in both these cases is different.

In the opposite limit $\lambda_{xx} \to 0$, or $\lambda_{xx} << |z_{VV}|, a_x z_{Vx}$, under the same condition $Sp\mathbf{Z} > 0$, Eq. (95) yields

$$\kappa_1 = \frac{\det \mathbf{Z}}{Sp\mathbf{Z}} = \frac{z_{VV}}{z_{VV} + a_x z_{Vx}}\lambda_{xx} \tag{97}$$

These results agree with the general rule that the nucleation rate is determined by the *slowest* kinetic process in the system [1, 9-11].

Similarly to nonisothermal effect in droplet nucleation [1, 11], we can characterize the deviation from the one-dimensional nucleation kinetics by considering the ratio

$$\varepsilon \equiv \frac{I}{I^{(1D)}} = \frac{\kappa_1}{z_{VV}} \tag{98}$$

Eq. (91) for $\lambda_{xx}$ allows us to determine some conditions for the above limits. It was mentioned above that $h_{xx}$ is large for a dilute precipitate ($x_* \sim 1$), which is the thermodynamic limit; however, this precipitate grows from the mother phase dilute with respect to component A, i.e. $x_0 << 1$. If, according to Eq. (88), $\xi_A << \xi_B$, then $\lambda_{xx} \sim h_{xx}\xi_A \sim x_0/(1-x_*)$, i.e. $\lambda_{xx}$ is determined by the relation of two small quantities which can be assumed of the same order of magnitude. This analysis implies that $D_0^A \sim D_0^B$. If $D_0^A >> D_0^B$ and therefore $\xi_A \sim \xi_B$, then $\lambda_{xx}$ is of the same order of magnitude as $h_{xx}$ and hence also large; so, the one-dimensional limit requires here the kinetic condition $D_0^A >> D_0^B$. This is only a qualitative reasoning; a more exact quantitative criterion can be obtained by numerical comparison of $\lambda_{xx}$ with $|z_{VV}|$ and $a_x z_{Vx}$. As was shown by the example of binary droplet nucleation [1], the two-dimensional process does not differ greatly from the one-dimensional one ($\varepsilon \sim 1$), even if $\lambda_{xx}$ and $|z_{VV}|$ are of the same order of magnitude; this means that the characteristic times of volume change and composition change are the same and the composition has a chance to adjust to the change in volume.

For a regular solution, the condition $\lambda_{xx} \rightarrow 0$ can be ensured by $\xi_i \rightarrow 0$; e.g., $\lambda_{xx} \sim \xi_A$, if $\xi_A \ll \xi_B$. According to Eq. (88), the latter condition can be fulfilled due to either $x_0 \rightarrow 0$ (the mother phase is dilute with respect to component A) or $D_0^A \ll D_0^B$. This case is similar to the nonisothermal limit in droplet nucleation [11, 12]; we have $\varepsilon \rightarrow 0$, and the problem is essentially two-dimensional.

In addition to the nucleation rate $I$, the matrices $\mathbf{H}$ and $\mathbf{Z}$ determine the steady state distribution function of nuclei $F_{st}(V,x)$. It allows us to calculate the mean steady state enrichment of nuclei with respect to one of the components [1]. This effect is a manifestation of two-dimensionality; it is more pronounced for small values of $\lambda_{xx}$. The nucleus composition does not have time to relax and its systematic deviation from $x_*$ occurs in the steady state. The similar effect is the mean steady state overheat of droplets in nonisothermal condensation [1, 11, 12].

## 5. Conclusion

The universal approach combining classical thermodynamics and the macroscopic kinetics of nucleus growth was applied here to the calculation of nucleation rates of precipitates in condensed binary solutions. Accordingly, the nucleation rate is expressed via thermodynamic ($T$, $x_0$, $\omega^\alpha$, etc.) and macroscopic kinetic ($D_0^i$, $\Gamma_i'$) parameters only. The macroscopic equations of nucleus growth, Eqs. (58a, b), have a simple form – the change in the number of $i$ atoms in unit time is proportional to the difference $(c_0^i - c_e^i)$ between the current concentration of $i$ atoms in the mother phase and their equilibrium concentration $c_e^i(R,x)$ over the nucleus of radius $R$ and composition $x$ (in the two-dimensional problem). This form is a result of application of the principle similar to detailed balancing in statistical mechanics: a nucleus loses atoms with the same frequency as in the state of equilibrium with the mother phase. In other words, the probability of losing an atom is determined by the nucleus state ($R,x$), rather than by the mother-phase state. While a nucleus grows ($R$ and $x$ change), the quantities $c_e^i(R,x)$ change according to thermodynamic Eqs. (79), (92), or other. The nucleus grows with respect to component $i$, while $c_0^i > c_e^i(R,x)$; the growth stops, when $c_e^i(R,x)$ becomes equal to $c_0^i$, and the nucleus dissolves with respect to $i$ at $c_0^i < c_e^i(R,x)$. So, these simple kinetic equations describe a complex evolution of a nucleus – the change both in size and composition.

It should be noted that the results of the present $(V,x)$ - theory can be reformulated in terms of the variables $(N_1, N_2)$ in the same way, as for a binary droplet in Ref. [1]. However, just the variable $x$ is natural for solving the given problem for the following reasons: (i) equations of equilibrium for a critical nucleus in Section 2 are formulated in terms of $x$; (ii) equations of nucleus growth are also written in terms of $x$; (iii) Eq. (62) for $\dot{x}$ allows us to reveal the parameter $\lambda_{xx}$ governing the nucleation kinetics. The work of a near-critical nucleus formation in the $(V, x)$-theory is a quadratic form with *diagonal* matrix.

The cases of both unary and binary precipitates were studied. As a particular case of binary precipitates, the nucleation of a precipitate of fixed composition (compound) was considered, which is the subject of Russell' theory [24]; this problem is solved here as a one-dimensional one. However, the most significant result of the present approach is the kinetics of nucleation of binary precipitates of a variable composition from non-ideal solutions, which is a two-dimensional problem. It is shown that the theory is consistent with Onsager's principle of symmetry of kinetic coefficients; also, the results are similar to those for a binary droplet nucleation [1], despite the fact that basic equations are different. The nucleation of precipitates of a fixed composition is also possible here as a one-dimensional limit of the theory at a large value of the kinetic parameter $\lambda_{xx}$; the composition rapidly relaxes to its critical value, and a nucleus growth with composition $x = x_*$ in the CNT approximation (beyond this approximation, $x$ relaxes to $x_{eq}(V)$ - see Appendix).

## Appendix: Surface effects in binary nucleation

### 1. Equations for the chemical potentials of bulk and surface phases

We consider the three-phase system consisting of new (bulk) phase $\alpha$, mother (bulk) phase $\beta$ and the surface layer between them which is phase $\sigma$ (Fig. 2); the surface of tension [27, 34] is employed as a dividing surface. This *finite-thickness layer method* (which is an alternative to Gibbs' one) was developed in detail for curved interfaces by Rusanov [27, 28]. The fundamental equations for all these phases are given in Ref. [12]; only some of them are needed for our purpose. Equation

$$a\,d\sigma + s^\sigma dT^\sigma - \upsilon^{\alpha\sigma} dP^\alpha - \upsilon^{\beta\sigma} dP^\beta + \sum_{i=A,B} x_i^\sigma d\mu_i^\sigma = 0 \qquad (A1)$$

is an analogue of Gibbs' adsorption equation; here $a = A/N^\sigma$, $s^\sigma = S^\sigma/N^\sigma$, $\upsilon^{\alpha\sigma} = V^{\alpha\sigma}/N^\sigma$, $\upsilon^{\beta\sigma} = V^{\beta\sigma}/N^\sigma$, $x_i^\sigma = N_i^\sigma/N^\sigma$ and $A$ is the nucleus surface area, $S^\sigma$ is the entropy of the surface layer, $N^\sigma$ is the total number of particles in it, $V^\sigma = V^{\alpha\sigma} + V^{\beta\sigma}$. It is seen that these specific quantities are the *mean* values for the surface layer, by definition.

Hereafter we put $x_A \equiv x$ and $x_B = 1 - x$ for a binary system, as well as denote

$$\dot{\mu}_A^\alpha \equiv \left(\frac{\partial \mu_A^\alpha}{\partial x^\alpha}\right)_{T^\alpha, P^\alpha} , \quad \dot{\mu}_B^\alpha \equiv \left(\frac{\partial \mu_B^\alpha}{\partial x^\alpha}\right)_{T^\alpha, P^\alpha}$$

and similarly for phases $\beta$ and $\sigma$. An equation for the chemical potential $\mu_i^\sigma(T^\sigma, P^\alpha, P^\beta, \sigma, x^\sigma)$ of component $i$ in the surface layer is [12]

$$d\mu_i^\sigma = -s_i^\sigma dT^\sigma + v_i^{\alpha\sigma} dP^\alpha + v_i^{\beta\sigma} dP^\beta - a_i d\sigma + \dot{\mu}_i^\sigma dx^\sigma \qquad (A2)$$

where $s_i^\sigma$, etc. are the partial molecular quantities,

$$s_i^\sigma = \left(\frac{\partial S^\sigma}{\partial N_i^\sigma}\right)_{T^\sigma, P^\alpha, P^\beta, \sigma, N_{j\neq i}^\sigma} \quad , \quad a_i = \left(\frac{\partial A}{\partial N_i^\sigma}\right)_{T^\sigma, P^\alpha, P^\beta, \sigma, N_{j\neq i}^\sigma} \quad , \text{ etc.}$$

For a bulk phase,

$$d\mu_i^\alpha = -s_i^\alpha dT^\alpha + v_i^\alpha dP^\alpha + \dot{\mu}_i^\alpha dx^\alpha \qquad (A3)$$

and the same for phase $\beta$. The isothermal-isobaric Gibbs-Duhem equation and Eq. (A1) for constant $T^\sigma$, $P^\alpha$, $P^\beta$, and $\sigma$ result in the following relations:

$$\dot{\mu}_B^\alpha = -\frac{x^\alpha}{1-x^\alpha} \dot{\mu}_A^\alpha \quad , \qquad \dot{\mu}_B^\sigma = -\frac{x^\sigma}{1-x^\sigma} \dot{\mu}_A^\sigma \qquad (A4)$$

In the state of full equilibrium, the equality $\mu_i^\alpha = \mu_i^\sigma = \mu_i^\beta$ holds [12]; its differential form

$$d\mu_i^\alpha = d\mu_i^\sigma = d\mu_i^\beta \qquad (A5)$$

is used below for deriving the needed relations, as well as equation

$$dP^\alpha = dP^\sigma + dP_L \qquad (A6)$$

Also $T^\alpha = T^\sigma = T^\beta \equiv T$, where $T$ is the common temperature of the system in equilibrium.

The complex consisting of phases $\alpha$ and $\sigma$ being in equilibrium with each other (but generally not in equilibrium with phase $\beta$) is the *density fluctuation* (DF) [35, 36] within phase $\beta$; its volume is by $V^{\beta\sigma}$ greater, than the nucleus volume $V$ bounded by the surface of tension (Fig.2).

## 1. Equilibrium of bulk phases

Equation $d\mu_i^\alpha = d\mu_i^\beta$ for $i = $ A and B results in the following system of equations:

$$\begin{cases} \dot{\mu}_A^\beta dx^\beta = (s_A^\beta - s_A^\alpha)dT + v_A^\alpha dP^\alpha - v_A^\beta dP^\beta + \dot{\mu}_A^\alpha dx^\alpha \\ \dot{\mu}_B^\beta dx^\beta = (s_B^\beta - s_B^\alpha)dT + v_B^\alpha dP^\alpha - v_B^\beta dP^\beta + \dot{\mu}_B^\alpha dx^\alpha \end{cases} \qquad (A7)$$

Expressing $\dot{\mu}_B^\alpha$ via $\dot{\mu}_A^\alpha$ and $\dot{\mu}_B^\beta$ via $\dot{\mu}_A^\beta$ according to Eq. (A4), we reduce this set to the following equation:

$$\Delta s_{(\alpha\beta)}dT + (v^\beta - \Delta v_{(\alpha\beta)})dP^\alpha - v^\beta dP^\beta = \frac{x^\alpha - x^\beta}{1-x^\alpha} \dot{\mu}_A^\alpha dx^\alpha \qquad (A8)$$

with

$$\Delta s_{(\alpha\beta)} \equiv s^\beta - s^\alpha - (x^\beta - x^\alpha)(s_A^\alpha - s_B^\alpha)$$

$$\Delta v_{(\alpha\beta)} \equiv v^\beta - v^\alpha - (x^\beta - x^\alpha)(v_A^\alpha - v_B^\alpha)$$

This equation gives the relationship between the composition of phase $\alpha$ and the system state parameters – temperature and pressures. For the flat interface, $P^\alpha = P^\beta$, it reduces to the equation derived by Van der Waals [37].

## 2. Relations between the compositions of coexisting phases

As is known, the composition of the nucleus surface layer differs from the composition of bulk phase $\alpha$, i.e. *adsorption* takes place. The need to incorporate the phenomenon of adsorption into the nucleation theory was noted by Wilemski [38, 39]. The above fundamental equations for a surface layer together with the equations of equilibrium allow deriving equations for the surface composition $x^\sigma$ at different conditions. Below, relations between $x^\sigma$ and $x^\alpha$, as well as between $x^\alpha$ and $x^\beta$, are derived.

Eq. (A7) is employed for deriving the dependence $x^\alpha(x^\beta)$. Substituting Eq. (A6) to this set of equations and then excluding $dP_L$ from it, we get an equation connecting $x^\alpha$, $x^\beta$, $T$, and $P^\beta$:

$$\frac{\upsilon^\alpha + (x^\beta - x^\alpha)(\upsilon_A^\alpha - \upsilon_B^\alpha)}{1 - x^\beta} \dot\mu_A^\beta dx^\beta = \left[\upsilon_B^\beta(s_A^\alpha - s_A^\alpha) - \upsilon_A^\alpha(s_B^\beta - s_B^\alpha)\right]dT$$

$$+ \left[\upsilon_A^\alpha \upsilon_B^\beta - \upsilon_B^\alpha \upsilon_A^\beta\right]dP^\beta + \frac{\upsilon^\alpha}{1 - x^\alpha} \dot\mu_A^\alpha dx^\alpha \tag{A9}$$

Isothermal-isobaric dependences are of the most practical interest. Thus, one obtains from this equation

$$\left(\frac{dx^\alpha}{dx^\beta}\right)_{T,P^\beta} = \frac{1 - x^\alpha}{1 - x^\beta}\left[1 + \frac{\upsilon_A^\alpha - \upsilon_B^\alpha}{\upsilon^\alpha}\left(x^\beta - x^\alpha\right)\right]\frac{\dot\mu_A^\beta}{\dot\mu_A^\alpha} \equiv \frac{1 - x^\alpha}{1 - x^\beta}\left[\frac{\upsilon_B^\alpha + (\upsilon_A^\alpha - \upsilon_B^\alpha)x^\beta}{\upsilon^\alpha}\right]\frac{\dot\mu_A^\beta}{\dot\mu_A^\alpha} \tag{A10}$$

where equation $\upsilon^\alpha = \upsilon_A^\alpha x^\alpha + \upsilon_B^\alpha(1 - x^\alpha)$ was utilized.

For ideal solutions, $\dot\mu_A^\alpha = kT / x^\alpha$ and $\dot\mu_A^\beta = kT / x^\beta$, this equation acquires the form

$$\left(\frac{dx^\alpha}{dx^\beta}\right)_{T,P^\beta} = \frac{x^\alpha(1 - x^\alpha)}{x^\beta(1 - x^\beta)}\left[\frac{\upsilon_B^\alpha + (\upsilon_A^\alpha - \upsilon_B^\alpha)x^\beta}{\upsilon^\alpha}\right] \tag{A11}$$

For deriving the relation between $x^\sigma$ and $x^\alpha$, we use $d\mu_i^\alpha = d\mu_i^\sigma$ with account for Eqs. (A3) and (A2):

$$\begin{cases} -s_A^\sigma dT + \upsilon_A^{\alpha\alpha}dP^\alpha + \upsilon_A^{\beta\alpha}dP^\beta - a_A d\sigma + \dot\mu_A^\sigma dx^\sigma = -s_A^\alpha dT + \upsilon_A^\alpha dP^\alpha + \dot\mu_A^\alpha dx^\alpha \\ -s_B^\sigma dT + \upsilon_B^{\alpha\alpha}dP^\alpha + \upsilon_B^{\beta\alpha}dP^\beta - a_B d\sigma + \dot\mu_B^\sigma dx^\sigma = -s_B^\alpha dT + \upsilon_B^\alpha dP^\alpha + \dot\mu_B^\alpha dx^\alpha \end{cases} \tag{A12}$$

Further, the following steps are done: (i) Eq. (A1) for $d\sigma$ with $d\mu_i^\alpha$ instead of $d\mu_i^\sigma$, as well as Eq. (A6) are substituted; (ii) Eq. (A4) for $\mu_i^\alpha$ is employed; (iii) $dP_L$ is excluded from the resulting set of equations and Eq. (A4) for $\mu_i^\sigma$ is employed. As a result, an equation connecting $x^\sigma$, $x^\alpha$, $T$, and $P^\beta$ is derived. The isothermal-isobaric equation is then obtained as follows:

$$\left(\frac{dx^{\sigma}}{dx^{\alpha}}\right)_{T,P^{\beta}} = \frac{1-x^{\sigma}}{1-x^{\alpha}}\left[\frac{a_B + (a_A - a_B)x^{\alpha}}{a}\right]\frac{\dot{\mu}_A^{\alpha}}{\dot{\mu}_A^{\sigma}} \equiv \frac{1-x^{\sigma}}{1-x^{\alpha}}\left[1 + \frac{(a_A - a_B)}{a}\left(x^{\alpha} - x^{\sigma}\right)\right]\frac{\dot{\mu}_A^{\alpha}}{\dot{\mu}_A^{\sigma}} \tag{A13}$$

where equation $a = a_A x^{\sigma} + a_B(1 - x^{\sigma})$ was utilized.

For ideal solutions in both phase $\alpha$ and surface layer,

$$\left(\frac{dx^{\sigma}}{dx^{\alpha}}\right)_{T,P^{\beta}} = \frac{x^{\sigma}(1-x^{\sigma})}{x^{\alpha}(1-x^{\alpha})}\left[\frac{a_B + (a_A - a_B)x^{\alpha}}{a}\right] \tag{A14}$$

It is seen that Eqs. (A13) and (A10) as well as (A14) and (A11) are quite similar. In the approximation of constant $v_i^{\alpha}$ (not depending on $x^{\alpha}$) and constant $a_i$ (not depending on $x^{\sigma}$), Eqs. (A11) and (A14) are equations with separated variables which are easily integrated.

Eq.(A13) allows us to determine the surface layer composition $x^{\sigma}$ for a given $x^{\alpha}$, i.e. to find the difference $(x^{\sigma} - x^{\alpha})$ of compositions of surface and bulk phases (adsorption). Of course, an equation for $(dx^{\sigma}/dx^{\beta})_{T,P^{\beta}}$ as well as other relations of interest can be derived in a similar way.

## 3. Dependence of surface tension on the new-phase composition

Eq. (A1) is basic for determining the dependences of surface tension on different state parameters of coexisting phases. We replace $d\mu_i^{\sigma}$ by $d\mu_i^{\alpha}$ and $d\mu_i^{\beta}$ in this equation and express $\dot{\mu}_B^{\alpha(\beta)}$ via $\dot{\mu}_A^{\alpha(\beta)}$, according to Eq. (A4); then Eq. (A7) for $\dot{\mu}_A^{\beta}dx^{\beta}$ and Eq. (A6) are employed. As a result, the following set of equations is obtained:

$$ad\sigma = \Delta s_{(\sigma\alpha)}dT - \left[\Delta v_{(\sigma\alpha)} + v^{\beta\sigma}\right]dP_L - \Delta v_{(\sigma\alpha)}dP^{\beta} + \frac{x^{\alpha} - x^{\sigma}}{1 - x^{\alpha}}\dot{\mu}_A^{\alpha}dx^{\alpha} \tag{A15}$$

$$(1-x^{\beta})ad\sigma = \left[\Delta s_{(\sigma\beta)}(1-x^{\beta}) + (x^{\beta} - x^{\sigma})(s_A^{\beta} - s_A^{\alpha})\right]dT + \left[(1-x^{\beta})v^{\alpha\sigma} + (x^{\beta} - x^{\sigma})v_A^{\alpha}\right]dP_L$$

$$+ \left[-\Delta v_{(\sigma\beta)}(1-x^{\beta}) + (x^{\beta} - x^{\sigma})(v_A^{\alpha} - v_A^{\beta})\right]dP^{\beta} + (x^{\beta} - x^{\sigma})\dot{\mu}_A^{\alpha}dx^{\alpha} \tag{A16}$$

with

$$\Delta s_{(\sigma\alpha)} \equiv s^{\alpha} - s^{\sigma} - (x^{\alpha} - x^{\sigma})(s_A^{\alpha} - s_B^{\alpha})$$

$$\Delta s_{(\sigma\beta)} \equiv s^{\beta} - s^{\sigma} - (x^{\alpha} - x^{\sigma})(s_A^{\beta} - s_B^{\beta})$$

$$\Delta v_{(\sigma\alpha)} \equiv v^{\alpha} - v^{\sigma} - (x^{\alpha} - x^{\sigma})(v_A^{\alpha} - v_B^{\alpha})$$

$$\Delta v_{(\sigma\beta)} \equiv v^{\beta} - v^{\sigma} - (x^{\alpha} - x^{\sigma})(v_A^{\alpha} - v_B^{\beta})$$

Excluding $dP_L$ from this set of equations, we get an equation connecting $\sigma$ with $x^{\alpha}$, $T$, and $P^{\beta}$. An equation for the isothermal-isobaric dependence $\sigma(x^{\alpha})_{T,P^{\beta}}$ has the following form:

$$(ad\sigma)_{T,P^{\beta}} = \frac{v^{\alpha}(x^{\beta} - x^{\sigma}) + v^{\alpha\sigma}(x^{\alpha} - x^{\beta})}{v_B^{\beta} + x^{\beta}(v_A^{\alpha} - v_B^{\alpha})}\frac{\dot{\mu}_A^{\alpha}}{1 - x^{\alpha}}dx^{\alpha} \tag{A17}$$

The similar equation obtained in Ref. [27] is reduced to Eq. (A17) after simple transformations. For integrating this equation, it has to be complemented by the functions $x^\beta(x^\alpha)_{T,P^\beta}$ and $x^\sigma(x^\alpha)_{T,P^\beta}$ which are determined from Eqs. (A10) and (A13).

Excluding $dP^\beta$ from Eqs. (A15) and (A16) and utilizing equation

$$dP_L = \frac{2}{R}d\sigma - \frac{2\sigma}{R^2}dR \tag{A18}$$

we get an equation connecting $\sigma$ with $x^\alpha$, $T$, and $R$ [27]

$$\left[a + \frac{2}{R}\left(v^{\beta\sigma} - \frac{\Delta v_{(\alpha\sigma)}}{\Delta v_{(\alpha\beta)}}v^\beta\right)\right]d\sigma = \left[\Delta s_{(\alpha\beta)}\frac{\Delta v_{(\alpha\sigma)}}{\Delta v_{(\alpha\beta)}} - \Delta s_{(\alpha\sigma)}\right]dT$$

$$+\frac{2\sigma}{R^2}\left(v^{\beta\sigma} - \frac{\Delta v_{(\alpha\sigma)}}{\Delta v_{(\alpha\beta)}}v^\beta\right)dR + \left[(x^\beta - x^\alpha)\frac{\Delta v_{(\alpha\sigma)}}{\Delta v_{(\alpha\beta)}} - (x^\sigma - x^\alpha)\right]\frac{\dot{\mu}_A^\alpha}{1 - x^\alpha}dx^\alpha \tag{A19}$$

where $\Delta v_{(\alpha\sigma)} = -\Delta v_{(\sigma\alpha)}$, $\Delta s_{(\alpha\sigma)} = -\Delta s_{(\sigma\alpha)}$. From here, an equation for $\sigma(x^\alpha)_{T,R}$ is obtained as follows:

$$\left[a + \frac{2}{R}\left(v^{\beta\sigma} - \frac{\Delta v_{(\alpha\sigma)}}{\Delta v_{(\alpha\beta)}}v^\beta\right)\right](d\sigma)_{T,R} = \left[(x^\beta - x^\alpha)\frac{\Delta v_{(\alpha\sigma)}}{\Delta v_{(\alpha\beta)}} - (x^\sigma - x^\alpha)\right]\frac{\dot{\mu}_A^\alpha}{1 - x^\alpha}dx^\alpha \tag{A20}$$

It is of interest to consider the dependence $\sigma(x^\alpha)_{T,R}$ following from the condition of the DF internal equilibrium only, i.e. from the condition of equilibrium between phases $\alpha$ and $\sigma$ at a fixed state of phase $\beta$, $(d\mu_i^\alpha = d\mu_i^\sigma)_{(\beta)}$. It is easily derived from Eq. (A15) which is just an equation for the internal equilibrium. Employing Eq. (A18) and then putting $dP^\beta = 0$ in Eq. (A15), we get

$$\left[a + \frac{2(\Delta v_{(\sigma\alpha)} + v^{\beta\sigma})}{R}\right](d\sigma)^{(DF)} = \Delta s_{(\sigma\alpha)}dT + \frac{2\sigma(\Delta v_{(\sigma\alpha)} + v^{\beta\sigma})}{R^2}dR + \frac{x^\alpha - x^\sigma}{1 - x^\alpha}\dot{\mu}_A^\alpha dx^\alpha \tag{A21}$$

from where

$$\left[a + \frac{2(\Delta v_{(\sigma\alpha)} + v^{\beta\sigma})}{R}\right](d\sigma)_{T,R}^{(DF)} = \frac{x^\alpha - x^\sigma}{1 - x^\alpha}\dot{\mu}_A^\alpha dx^\alpha \tag{A22}$$

For a liquid droplet ($\alpha$) in vapor ($\beta$), Eqs. (A19) and (A20) are simplified due to the condition $\Delta v_{(\alpha\beta)} \approx v^\beta >> \Delta v_{(\alpha\sigma)}$ and go into Eqs. (A21) and (A22), respectively. So, for a liquid binary droplet in vapor, an equation for $\sigma(x^\alpha)_{T,R}$ derived from the condition of full equilibrium coincides with that derived from the condition of the DF internal equilibrium only.

It is seen from Eq. (A22) that the surface tension depends on $x^\alpha$ due to *adsorption*, or the difference in compositions $x^\alpha$ and $x^\sigma$. For integrating this equation, the dependence $x^\sigma(x^\alpha)_{T,R}^{(DF)}$ is needed. It is obtained from Eq. (A12) which is just an internal equilibrium equation. Substituting $d\sigma$ from Eq. (A21) in Eq. (A12), as well as utilizing Eq. (A18) and putting $dP^\beta = 0$, we get from any equation of this set

$$\left[a+\frac{2(\Delta\upsilon_{(\sigma\alpha)}+\upsilon^{\beta\sigma})}{R}\right]\left(\frac{dx^{\sigma}}{dx^{\alpha}}\right)_{T,R}^{(DF)}=\frac{1-x^{\sigma}}{1-x^{\alpha}}\Big\{a_B+(a_A-a_B)x^{\alpha}$$

$$+\frac{2}{R}\Big[(\upsilon^{\alpha}-\upsilon^{\alpha\sigma})-(x^{\alpha}-x^{\sigma})(\upsilon_A^{\alpha\sigma}-\upsilon_B^{\alpha\sigma})\Big]\Big\}\frac{\dot{\mu}_A^{\alpha}}{\dot{\mu}_A^{\sigma}} \tag{A23}$$

In the limit of planar interface, this equation goes into Eq. (A13). $T$ in Eqs. (A21)-(A23) is the DF temperature: $T\equiv T^{\alpha}=T^{\sigma}$.

## 4. Surface effects on the work of binary nucleus formation

The second differential of the work with the surface layer contribution was calculated in Ref. [12]; for a binary nucleus, it has the following form:

$$(d^2W)_{(\beta)}^{*}\equiv-\frac{P_L^{*}}{3V_*}(dV)^2+H^{\alpha}+H^{\sigma}$$

$$H^{\alpha}=\sum_{i=A,B}\Big\{N_{i*}^{\alpha}\big[ds_i^{\alpha}dT-d\upsilon_i^{\alpha}dP^{\alpha}\big]_*+\dot{\mu}_{i*}^{\alpha}dx^{\alpha}dN_i^{\alpha}\Big\}$$

$$H^{\sigma}=\sum_{i=A,B}\Big\{N_{i*}^{\sigma}\big[ds_i^{\sigma}dT-d\upsilon_i^{\alpha\sigma}dP^{\alpha}\big]_*+\dot{\mu}_{i*}^{\sigma}dx^{\sigma}dN_i^{\sigma}\Big\}+\sum_{i=A,B}N_{i*}^{\sigma}da_id\sigma \tag{A24}$$

where $H^{\sigma}$ is just the mentioned surface layer contribution. $H^{\alpha}$ and $H^{\sigma}$ are the positive definite quadratic forms of stable variables for phases $\alpha$ and $\sigma$; only $H^{\alpha}$ enters this equation in the CNT approximation.

The matrix of the quadratic form $H^{\alpha}$ for an incompressible droplet was found in Ref. [1] as

$$\mathbf{H}^{\alpha}=\begin{pmatrix}N_*^{\alpha}\dfrac{\dot{\mu}_{A*}^{\alpha}}{1-x_*^{\alpha}} & 0\\[3mm] 0 & \dfrac{C_{V*}^{\alpha}}{T^{\beta}}\end{pmatrix} \tag{A25}$$

with $N_*^{\alpha}=N_{A*}^{\alpha}+N_{B*}^{\alpha}$; $C_{V*}^{\alpha}$ is the heat capacity of the critical droplet. The same equation with the replacement of superscript $\alpha$ by $\sigma$ holds for the expression in braces for $H^{\sigma}$, Eq. (A24), whereas the last term is represented as follows:

$$d\sigma=\left(\frac{\partial\sigma}{\partial R}\right)_{T,x^{\alpha}}\frac{dR}{dV}dV+\left(\frac{\partial\sigma}{\partial x^{\alpha}}\right)_{R,T}dx^{\alpha}+\left(\frac{\partial\sigma}{\partial T}\right)_{R,x^{\alpha}}dT \tag{A26}$$

$$\sum_{i=A,B}N_i^{\sigma}da_i=dA-\sum_{i=A,B}a_idN_i^{\sigma}=\left[\frac{2}{R}-\sum_{i=A,B}a_i\frac{dN_i^{\sigma}}{dV}\right]dV \tag{A27}$$

$$\sum_{i=A,B}N_{i*}^{\sigma}da_id\sigma=\frac{2\chi}{3V_*}\left(\frac{\partial\sigma}{\partial R}\right)_{T,x^{\alpha}}^{*}(dV)^2+\frac{2\chi}{R_*}\left(\frac{\partial\sigma}{\partial x^{\alpha}}\right)_{R,T}^{*}dVdx^{\alpha}+\frac{2\chi}{R_*}\left(\frac{\partial\sigma}{\partial T}\right)_{R,x^{\alpha}}^{*}dVdT \tag{A28}$$

where

$$\chi \equiv 1 - \frac{R_*}{2} \sum_{i=A,B} a_i \left( \frac{dN_i^\sigma}{dV} \right)_*$$

Droplet temperature is an important variable; it allows us to take into account nonisothermal effects in condensation [11, 12]. However, the heat conductivity in condensed matter is much higher, than in a vapor; thus, the temperature is not required as a variable of nucleus description and omitted below (it falls out from consideration as a result of the corresponding kinetic limit). From the above equations, the matrix of the work of binary nucleus formation in a condensed state is

$$\mathbf{H} = \begin{pmatrix} -\dfrac{1}{3V_*} \left[ P_L^* - 2\chi \left( \dfrac{\partial \sigma}{\partial R} \right)_{T,x^\alpha}^* \right] & \dfrac{\chi}{R_*} \left( \dfrac{\partial \sigma}{\partial x^\alpha} \right)_{R,T}^* \\[3mm] \dfrac{\chi}{R_*} \left( \dfrac{\partial \sigma}{\partial T} \right)_{R,x^\alpha}^* & N_*^\alpha \dfrac{\dot\mu_{A*}^\alpha}{1-x_*^\alpha} + N_*^\sigma \dfrac{\dot\mu_{A*}^\sigma}{1-x_*^\sigma} \left( \dfrac{dx^\sigma}{dx^\alpha} \right)_*^2 \end{pmatrix} \quad (A29)$$

Eq. (A24) for $(d^2W)^*_{(\beta)}$ was derived at a *fixed* state of phase $\beta$ [12], which is marked by the subscript; it was assumed that the mother-phase state does not change upon the nucleus formation. Therefore, equations derived above from the condition of DF internal equilibrium at a fixed state of phase $\beta$ are employed for determining matrix $\mathbf{H}$ elements. Specifically: (i) equations of Section 2 determines $R_*$ and $x_*^\alpha$ for a given state (supersaturation) of phase $\beta$; (ii) at the given $R_*$ and $T = T^\beta$, $x^\sigma$ is a function of $x^\alpha$ which is determined from Eq. (A23); the derivative $dx^\sigma / dx^\alpha$ is taken from this equation with $R = R_*$ also. (iii) The derivatives of surface tension in Eq. (A29) are determined by Eq. (A21).

When the nucleus is so small that phase $\alpha$ is absent, $N_*^\alpha \to 0$, $V = V^{\alpha\sigma}$, only the surface parameters remain in Eq. (A29); the set of variables ($V$, $x^\sigma$) is more convenient in this case for considering the kinetics of nucleus evolution. The element $h_{xx}$ in these variables has the form

$$h_{xx} = N_*^\alpha \frac{\dot\mu_{A*}^\alpha}{1-x_*^\alpha} \left( \frac{dx^\alpha}{dx^\sigma} \right)_*^2 + N_*^\sigma \frac{\dot\mu_{A*}^\sigma}{1-x_*^\sigma}$$

and the first summand vanishes together with $N_*^\alpha$. Thus, the matrix $\mathbf{H}$ becomes as follows:

$$\mathbf{H} = \begin{pmatrix} -\dfrac{1}{3V_*} \left[ P_L^* - 2\chi \left( \dfrac{\partial \sigma}{\partial R} \right)_{T,x^\alpha}^* \right] & \dfrac{\chi}{R_*} \left( \dfrac{\partial \sigma}{\partial x^\alpha} \right)_{R,T}^* \left( \dfrac{dx^\alpha}{dx^\sigma} \right)_* \\[3mm] \dfrac{\chi}{R_*} \left( \dfrac{\partial \sigma}{\partial x^\alpha} \right)_{R,T}^* \left( \dfrac{dx^\alpha}{dx^\sigma} \right)_* & N_*^\sigma \dfrac{\dot\mu_{1*}^\sigma}{1-x_*^\sigma} \end{pmatrix} \quad (A30)$$

The factor $\chi$ acquires the form

$$\chi \equiv 1 - \frac{R_*}{2} \sum_{i=A,B} \frac{a_i}{v_i^{\alpha\sigma}} \quad (A31)$$

From inequality

$$\sum_{i=A,B} \frac{a_i}{v_i^{\alpha\sigma}} - \frac{a}{v^{\alpha\sigma}} = \frac{a_B (v_A^{\alpha\sigma})^2 x^\sigma + a_A (v_B^{\alpha\sigma})^2 (1-x^\sigma)}{v_A^{\alpha\sigma} v_B^{\alpha\sigma} v^{\alpha\sigma}} > 0 \quad (A32)$$

it follows that

$$\frac{R}{2}\sum_{i=A,B}\frac{a_i}{\upsilon_i^{\alpha\sigma}} > \frac{Ra}{2\upsilon^{\alpha\sigma}} = \frac{3}{2} \quad \text{and} \quad \chi < -\frac{1}{2} \tag{A33}$$

i. e. $\chi < 0$, which is an important property [12].

Comparing Eq. (A30) to the CNT matrix $\mathbf{H}$, Eq. (48), we see that the dependence of surface tension on radius changes the element $h_{VV}$ (the nucleation barrier curvature), whereas the dependence on $x^\alpha$ yields the off-diagonal elements $h_{Vx}$. The presence of the derivative $dx^\sigma/dx^\alpha$ in Eqs. (A29) and (A30) shows that taking into account the adsorption phenomenon is naturally required, when the nucleation work is written with the surface term, whereas this is not the case for the CNT nucleation work, Eq. (48).

The quadratic form with the matrix $\mathbf{H}$, Eq. (A29), can be identically transformed as follows:

$$H(V,x) = h_{VV}(V-V_*)^2 + 2h_{Vx}(V-V_*)(x-x_*) + h_{xx}(x-x_*)^2$$

$$\equiv \frac{\det \mathbf{H}}{h_{xx}}(V-V_*)^2 + h_{xx}(x-x_{eq})^2 = \frac{\det \mathbf{H}}{h_{xx}}(V-V_*)^2 + h_{xx}(x'-x_*)^2 \tag{A34}$$

$$x_{eq}(V) = x_* - \frac{h_{Vx}}{h_{xx}}(V-V_*), \qquad x' = x + \frac{h_{Vx}}{h_{xx}}(V-V_*) \tag{A35}$$

Such transformation with respect to arbitrary stable variables was used in Ref. [8] for normalizing the equilibrium distribution function. The quantity $x_{eq}$ is a solution of equation $\partial H/\partial x = 0$, or $\partial W/\partial x = 0$; this fact together with the form of Eq. (A34) leads to the conclusion that $x_{eq}$ plays the role of *equilibrium composition* for a *noncritical* nucleus. So, $x_{eq} = x_*$ for noncritical nuclei holds only in the CNT approximation, where the dependence of surface tension on composition is absent, $h_{Vx} = 0$. While the critical nucleus composition fluctuates around $x_*$, the noncritical nucleus composition fluctuates around $x_{eq}$ with the same rms.

The same Eqs. (A34) and (A45) with replacement $x \to T$ hold for a unary droplet nucleation due to the dependence of surface tension on droplet temperature $T$ [12]. The presence of the off-diagonal elements $h_{Vx}$ does not allow us to get Eq. (64) for $a_x$ and hence to write Eq. (62). In order to solve this problem, we must go to the new variables, $(V,x) \to (V,x')$, where the matrix $\mathbf{H}'$ is diagonal. This procedure with respect to temperature is performed in Ref. [12]; so the resulting equations from this work can be employed here with replacement $T \to x$. The transition matrix is

$$\mathbf{C} = \begin{pmatrix} 1 & 0 \\ -h_{Vx}/h_{xx} & 1 \end{pmatrix}, \qquad \begin{pmatrix} V-V_* \\ x-x_* \end{pmatrix} = \mathbf{C}\begin{pmatrix} V-V_* \\ x'-x_* \end{pmatrix} \tag{A36}$$

The matrices $\mathbf{H}$ and $\mathbf{Z}$ are transformed according to equations $\mathbf{H}' = \mathbf{C}^{\mathrm{T}}\mathbf{H}\mathbf{C}$, $\mathbf{Z}' = \mathbf{C}^{-1}\mathbf{Z}\mathbf{C}$ and acquire the form

$$\mathbf{H}' = \begin{pmatrix} \dfrac{\det \mathbf{H}}{h_{xx}} & 0 \\ 0 & h_{xx} \end{pmatrix}, \quad \mathbf{Z}' = \begin{pmatrix} z_{VV} - \dfrac{h_{Vx}}{h_{xx}} z_{Vx} & z_{Vx} \\ \dfrac{\det \mathbf{H}}{h_{xx}^2} z_{Vx} & \dfrac{h_{Vx}}{h_{xx}} z_{Vx} + z_{xx} \end{pmatrix} \tag{A37}$$

As the matrix $\mathbf{H}'$ is diagonal now, we can write

$$\dot{x}' = a_x' \dot{V} - \lambda_{xx}'(x' - x_*) \tag{A38}$$

$$z_{xV}' = a_x' z_{VV}', \quad z_{xx}' = \lambda_{xx}' + a_x' z_{Vx}' \tag{A39}$$

and equation for $a_x'$ has the form of Eq. (64),

$$a_x' = \frac{z_{Vx}'}{z_{VV}'} \frac{h_{VV}'}{h_{xx}'} \tag{A40}$$

Substituting the elements from Eq. (A37), we get

$$a_x' = \left(1 - \frac{h_{Vx}^2}{h_{VV} h_{xx}}\right)\left(1 - \frac{h_{Vx}}{h_{VV}} a_x^0\right)^{-1} a_x^0, \quad a_x^0 \equiv \frac{z_{Vx}}{z_{VV}} \frac{h_{VV}}{h_{xx}} \tag{A41}$$

Substituting the obtained $a_x'$ in Eq. (A38) and returning to $x$, according to Eq. (A35), we have

$$\dot{x} = \left(1 - \frac{h_{Vx}}{h_{VV}} a_x^0\right)^{-1}\left[a_x^0 - \frac{h_{Vx}}{h_{xx}}\right]\dot{V} - \lambda_{xx}'(x - x_{eq}) \tag{A42}$$

The quantity $\lambda_{xx}'$ is obtained from Eq. (A39): $\lambda_{xx}' = z_{xx}' - a_x' z_{Vx}'$. Taking $z_{xx}'$ and $z_{Vx}'$ from Eq. (A37), we get after simple transformations

$$\lambda_{xx}' \equiv \lambda_{xx} = z_{xx} - \left(a_x^0 - \frac{h_{Vx}}{h_{xx}}\right)\left(1 - \frac{h_{Vx}}{h_{VV}} a_x^0\right)^{-1} z_{Vx} \tag{A43}$$

This is the form of equations for $\dot{x}$ and $\lambda_{xx}$ in the case of composition-dependent surface tension. The form of Eq. (A42) confirms the meaning of $x_{eq}$ as the equilibrium composition for the given $V$; the composition $x$ relaxes to $x_{eq}$, rather than to $x_*$. In the CNT approximation, $h_{Vx} = 0$, Eq. (A42) acquires the form of Eq. (62), and $\lambda_{xx}$ is transformed to its CNT value.

Finally, it should be noted that the equations of equilibrium of Section 2 for a binary precipitate must be rederived for the case of composition-dependent surface tension. If they have the same form, but with $\sigma = \sigma(R, x^\alpha)$ (as for a binary droplet [1]), nevertheless, the elements $z_{ik}$ in Eqs. (A41)-(A43) are not the CNT elements given by Eq. (89); they include the terms with derivatives $(\partial \sigma / \partial R)_*$ and $(\partial \sigma / \partial x^\alpha)_*$. The surface parameters entering the above equations can be estimated within statistical mechanics, the density functional theory [40-43], or obtained from computer simulations.

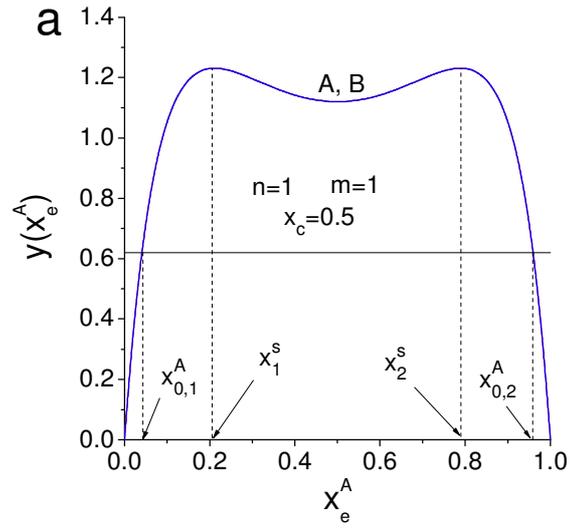

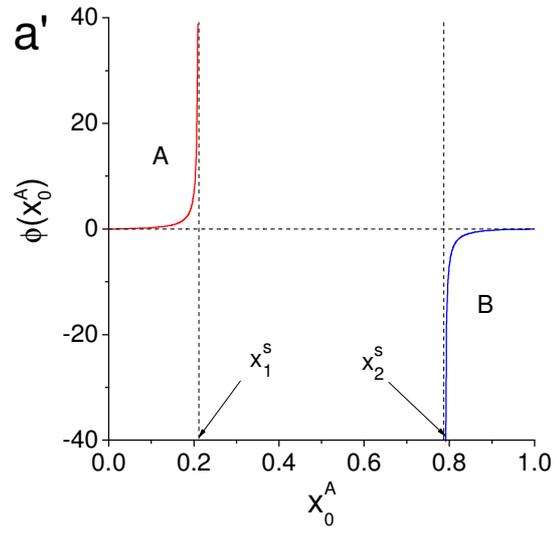

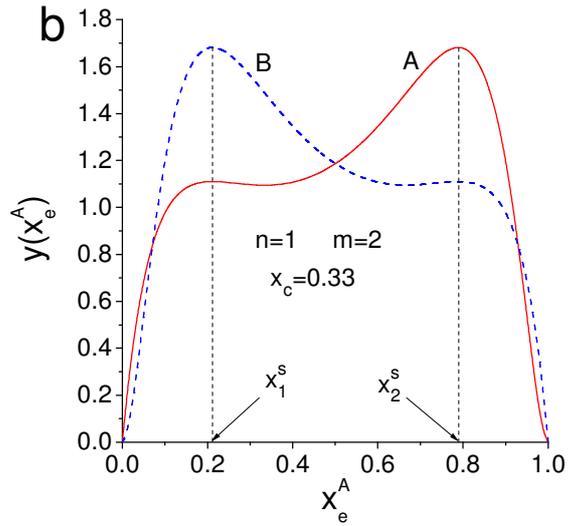

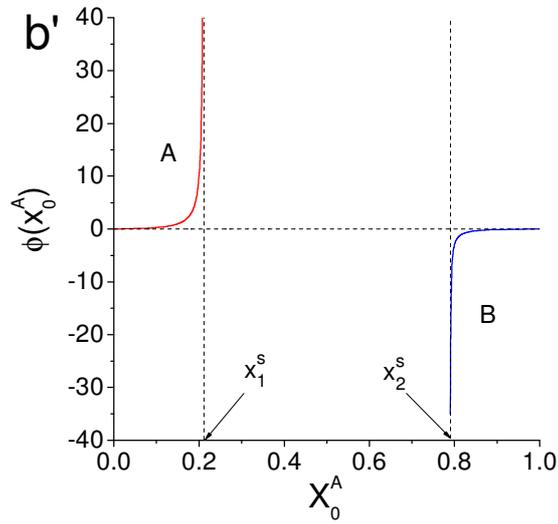

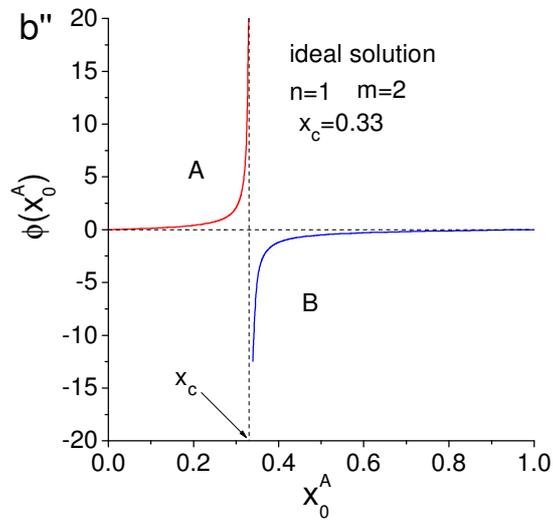

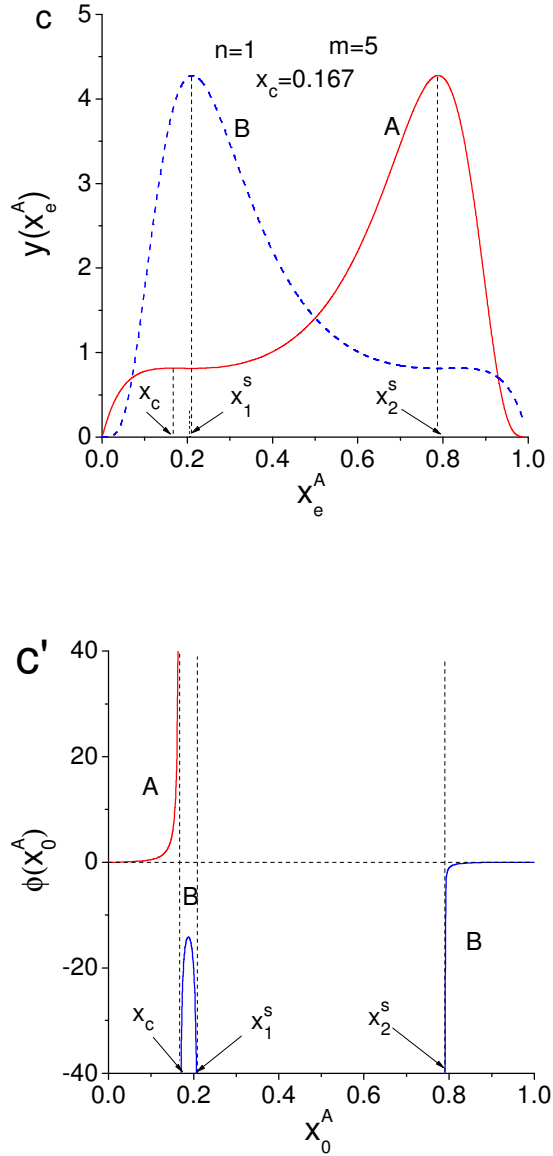

Fig. 1. The LHS of Eq. (73), $y(x_e^A)$ (solid) and $y(x_e^B)$ (dashed), for $\omega = 3$, $x_1^s = 0.21$, $x_2^s = 0.79$, and different pairs $(n, m)$ shown in figures (a), (b) and (c). The corresponding derivative $(dx_e^A / dP_L)_*$ - the function $\varphi(x_0^A)$ in Eq. (74a) – is shown in figures (a'), (b') and (c'). The symbols A and B show here the regions, where the corresponding component is a solute and determines the nucleation kinetics. Fig. (b'') shows the function $\varphi(x_0^A)$ for ideal solution, $\omega = 0$. Fig. (a) also demonstrates graphical solving Eq. (73): the straight line represents the value of the RHS of Eq. (73) for some critical radius $R_*$; it intersects the curve (LHS) at two points $x_{0,1}^A$ and $x_{0,2}^A$ corresponding to the given $R_*$.

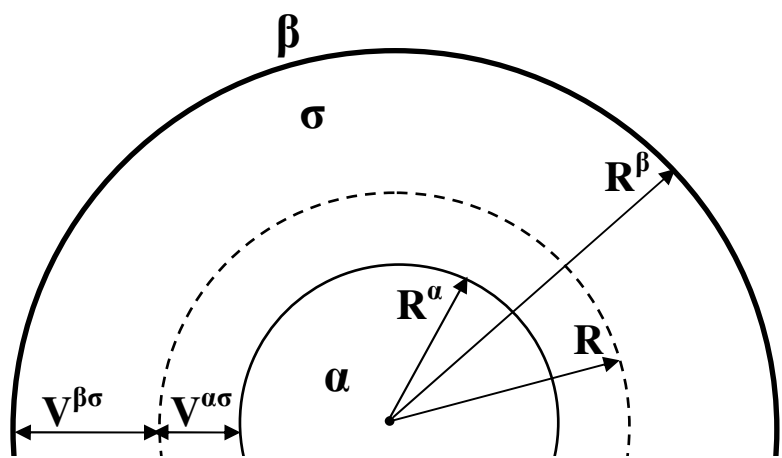

Fig. 2. Phases $\alpha$, $\sigma$, and $\beta$. The density fluctuation (DF) is bounded by bold line; the surface of tension is shown by dashed line.